\documentclass[aps,superscriptaddress,twocolumn,reprint]{revtex4-2}

\usepackage{times}
\usepackage{graphicx}
\usepackage{amsmath}
\usepackage{amsfonts}
\usepackage{amsmath}
\usepackage{amssymb}
\usepackage{amsthm}
\usepackage{bm}
\usepackage{amsopn}
\usepackage{xspace}
\usepackage{array}
\usepackage{placeins}
\usepackage{geometry}
\usepackage{comment}
\usepackage{subcaption}
\usepackage{cases}
\usepackage[outdir=./]{epstopdf}
\usepackage[dvipsnames]{xcolor}
\usepackage{tikz}
\usepackage{3dplot} %requires 3dplot.sty to be in same directory, or in your LaTeX installation
\usepackage{pgfplots}
\usetikzlibrary{spy}
\usepackage{xcolor}
\usepackage{makecell}

\newgeometry{right=1.5cm,left=1.5cm,top=1.8cm,bottom=1.8cm}

\newdimen\XCoord
\newdimen\YCoord
\usetikzlibrary{arrows.meta,positioning,calc,decorations.markings}

\begin{document}
%\iffalse
\author{C. Livi}
\affiliation{Fluids and Flows group and J.M. Burgers Centre for Fluid Dynamics, Department of Applied Physics, Eindhoven University of Technology, P.O. Box 513, 5600MB, Eindhoven, The Netherlands}
\author{G. Di Staso}
\affiliation{Fluids and Flows group and J.M. Burgers Centre for Fluid Dynamics, Department of Applied Physics, Eindhoven University of Technology, P.O. Box 513, 5600MB, Eindhoven, The Netherlands}
\affiliation{FLOW Matters Consultancy B.V., Groene Loper 5, 5612AE, Eindhoven, The Netherlands}
\author{{H. J. H.} Clercx}
\affiliation{Fluids and Flows group and J.M. Burgers Centre for Fluid Dynamics, Department of Applied Physics, Eindhoven University of Technology, P.O. Box 513, 5600MB, Eindhoven, The Netherlands}
\author{F. Toschi}
\affiliation{Fluids and Flows group and J.M. Burgers Centre for Fluid Dynamics, Department of Applied Physics, Eindhoven University of Technology, P.O. Box 513, 5600MB, Eindhoven, The Netherlands}
\begin{abstract}

The importance of accurately capturing two-way coupled interactions between particles with complex shapes and rarefied gas flows is rapidly rising in different practical applications such as aerospace industry and semiconductor manufacturing. The transport of particles in these conditions is often modelled via an Euler-Lagrangian Point-Particles approach, where rarefaction effects are included through the phenomenological Cunningham corrections on the drag force experienced by the particles. In Point-Particles  approaches, any explicit relation to the finite size of the particles, shape, orientation and momentum accommodation coefficient is typically neglected. In this work we aim to cover this gap by deriving, from fully-resolved DSMC simulations, heuristic models for the drag force acting on ellipsoidal particles with different aspect ratios. We include in the models the capability to predict effects related to gas-surface interactions via the tangential momentum accommodation coefficient (TMAC). The derived models can be used as corrections (to include shape, orientation and TMAC effects) in standard Euler-Lagrangian Point Particles simulations in rarefied gas flows. Additionally, we show that the obtained drag corrections, formally valid for unbounded gas flows, can potentially be applied also in cases where the particle moves in proximity to a solid wall. We do so by investigating near-wall effects on the drag of a prolate ellipsoidal particle.  Due to confinement effects, the drag increases when compared to the unbounded case, but such effects are typically negligible  for large $Kn$ also in cases in which the particle is in contact with the solid wall.

\end{abstract}
\title{Modelling drag coefficients of ellipsoidal particles in rarefied flow conditions }
\maketitle
\section{Introduction}

Micro and nanoparticles play an important role in a large variety of fields including aerospace industry \cite{jet1}, drug delivery \cite{lung} and contamination control in high-tech mechanical systems \cite{lito,immersionlito}. Such particles are produced and suspended in fluids during machine operations, components  handling, material processing, and by unintentional and/or undesirable release into the working environment. Often the suspending fluid is a gas in a low pressure environment, leading to non-negligible rarefaction effects. \\
In many numerical investigations, involving the transport of micro- and nano-sized particles in gas flows, Eulerian-Lagrangian simulations are employed \cite{xuzheng,shen,zhang,abduali,kaizhang,jet2,lungeuler}. In this approach, the flow field is evaluated on Eulerian grids, while spherical particles are modeled as Lagrangian points whose positions and velocities are evolved in time and rarefaction effects are included through the Cunningham correction \cite{cunningham}, which represents a rarefaction correction to the Stokes drag experienced by spherical particles.\\
While the Cunningham correction is widely used to model the drag force on spherical bodies, for non-spherical particles any explicit relation to their finite size and shape is neglected. Moreover, also when only spherical particles are considered, the Cunningham correction does not include a dependence on the momentum accommodation coefficient, i.e. the relation with the type of reflections that the gas molecules undergo when they hit the solid surface of the particle. From the pioneering work of Millikan \cite{millikan1, millikan2}, in fact, it is assumed that a large majority of gas-surface reflections is diffusive. This assumption has been later verified by Buckley et al. \cite{buckley}, who found that a value for the tangential accommodation coefficient of $\sigma=0.809$ described Millikan's results with good accuracy. If smaller particles, such as nano-particles, are considered, however, a larger fraction of specular reflections can appear \cite{wang}, in particular for particles with a diameter smaller than $20$nm, leading to a reduced accuracy of the Cunningham correction.\\
\begin{figure}[h!]
\vspace{0.2cm}

\centering
\scalebox{0.85}{
\vspace{0.1cm}

\begin{tikzpicture}[>=stealth]

    % Draw ellipse
      \begin{scope}[shift={(0,0)}, rotate around={-45:(0,0cm)},myVeryThick/.style={line width=1.2pt}]
           \draw [myVeryThick] (0,0) ellipse (1cm and 0.5 cm);
           \fill[gray!40] (0,0) ellipse (1cm and 0.5 cm);
           \draw[myVeryThick,-] (0,0) -- (-1,0cm);
            \draw[color=black,myVeryThick,<->] (0.05cm,0cm) -- (1cm,0);   
         \node[anchor=west,inner sep=0] at (1.1cm,0cm) {\Large{$a$}};
            \draw[color=black,myVeryThick,<->] (0.0cm,0.05cm) -- (0cm,0.5);   
         \node[anchor=south,inner sep=0] at (0cm,0.65cm) {\Large{$b$}};
      \end{scope}
      
    % angle of attack  
       \begin{scope}[shift={(0,0)}, myVeryThick/.style={line width=1.2pt}]
       	 \draw[myVeryThick,-] (0,0) -- (-0.8cm,0);
     	 \draw[color=black,myVeryThick] (-0.35cm,0.35cm) arc (135:180:0.5cm);   
         \node[anchor=east,inner sep=0] at (-0.85cm,0.32cm) {\Large{$\Phi$}};
     	 \draw[color=black,myVeryThick,->] (-2.8,0) -- (-1.9cm,0cm);  
     	   \draw[color=black,myVeryThick,->] (-2.8,0.8) -- (-1.9cm,0.8cm);  
     	   \draw[color=black,myVeryThick,->] (-2.8,-0.8) -- (-1.9cm,-0.8cm);  

         \node[anchor=south,inner sep=0] at (-2.2cm,0.2cm) {\Large{$U_0$}};         
         
       \end{scope}
    % sim box
       \begin{scope}[myVeryThick/.style={line width=2pt}]
      	 \draw[myVeryThick,dashed] (-3cm,-3cm) rectangle (3cm,3cm);
                \draw[color=black,line width=1.2pt,<->] (-2.9cm,2.7cm) -- (2.9cm,2.7cm);   
         \node[anchor=north,inner sep=0] at (0cm,2.6cm) {\Large{$L$}};

       \end{scope}

    % axes annotations 
       \begin{scope}[shift={(-2.2cm,-2.2cm)},myVeryThick/.style={line width=1pt}]
      	       	 \draw[myVeryThick,->] (0,0) -- (0.5cm,0);
      	       	 \draw[myVeryThick,->] (0,0) -- (0,0.5cm);
      	       	 \draw[myVeryThick,->] (0,0) circle (0.1);
                \node[anchor=west,inner sep=0]   at (0.6,0) {\Large{$\hat{x}$}};
                \node[anchor=south,inner sep=0] at (0,0.6) {\Large{$\hat{y}$}};
                \node[anchor=north,inner sep=0] at (-0.12,-0.12) {\Large{$\hat{z}$}};

       \end{scope}

\end{tikzpicture}
}
\caption{\small{Sketch of the simulation setup: an ellipsoidal particle with aspect ratio $a/b$ is immersed in a uniform gas flow with ambient velocity $U_0$. The particle is at the center of a cubic simulation box of size $L=20a$, at which free streaming boundary conditions are applied on the faces orthogonal to $\hat{x}$, and periodic boundary conditions elsewhere. We perform simulations varying the angle of attack, $\Phi$, the Knudsen number, $Kn$ (by varying the pressure and density of the gas), the aspect ratio, $a/b$, and the tangential momentum accommodation coefficient, $\sigma$. }}
\label{fig:sk_1}
\end{figure}
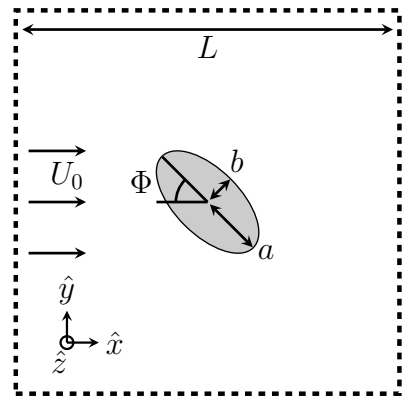
By extending our previous work \cite{livi}, in this study we aim to address these aspects by deriving accurate heuristic predictive models for the drag experienced by ellipsoidal particles with different aspect ratio, orientation and accommodation coefficient, immersed in a uniform rarefied gas flow. Such models are derived by fitting data from fully-resolved DSMC simulations, performed using our in-house code (extensively described in \cite{distaso1,distaso2,distaso_thesis}). One important aspect in this code concerns the two-way coupling between the gas molecules and the solid particle: In few words, the collision points between gas molecules and the surface of the solid particles are computed exactly through the use of a ray-sphere intersection method modified to include ellipsoidal particles, and the total force acting on the particle is computed via a momentum-exchange approach (for details, see \cite{livi}).\\
The core of our approach lays on the observation that the sine-squared drag law, firstly introduced by Happel and Brenner for the continuum and low-Reynolds number regime \cite {happel} (later extended by Sanjeevi \textit{et al.} to the high-Reynolds regime \cite{sanjeevi1,sanjeevi2}), is also valid in the low-Reynolds number, rarefied gas flow case. The sine-squared drag law, given by
\begin{align}
C_D(\Phi) = C_{D,0^\circ} + (C_{D,90^\circ} - C_{D,0^\circ} )\sin^2\Phi,
\label{eq:sine_squared_f}
\end{align}
states that the drag coefficient (and, thus, the drag force) of an arbitrary shaped particle at a given orientation $\Phi$ with respect to the uniform flow, see Fig. \ref{fig:sk_1}, can be fully characterized by its values at $0^\circ$ and $90^\circ$. In this way, it is sufficient to obtain predictive models able to reproduce the drag at the orientations of $0^\circ$ and $90^\circ$ and use Eq. (\ref{eq:sine_squared_f}) to extend the prediction in the whole range of $\Phi$.\\
We perform DSMC simulations at finite Knudsen number ($2\leq Kn \leq 10$) to derive the predictive models as analytical functions of $Kn,\ \Phi$ and $\sigma$, for different particle aspect ratios. We show that the predictions of the models are in good agreement with DSMC data that is not used during the fitting process, also when data outside of the fitting range is considered. The models derived in this work can be used to greatly improve Euler-Lagrangian Point Particles simulations by providing corrections to the drag coefficient of the particle that include effects with regards to shape, rarefaction, orientation and tangential accommodation coefficient.\\
While these results are formally valid only in unbounded fluids, in the last part of the paper we show that the drag corrections obtained with the proposed approach can be safely employed also in the non-ideal cases where the unbounded condition is not preserved, for example in cases when domain walls are present. Through a minor modification of the simulation setup, we investigate the near-wall effects on the drag force experienced by a prolate ellipsoidal particle (we focus on the prolate case to restrict the number of parameters in play) located in the vicinity of a solid wall for different orientations, distances from the wall and rarefaction levels.  We show that the $\Phi$-dependence of the drag force exhibits deviation with respect to the sine-squared drag law of the unbounded case, but such effects are negligible for $Kn\geq 2$ and they vanish quickly as soon as the particle distance with the wall increases. Since we aim to apply the drag corrections derived in this work for large Knudsen number and in cases where the particles spend on average the majority of the time far away from the walls, we can conclude that near-wall effects can be neglected in most of the cases of interest.\\
The rest of the paper is structured as follows: in Section \ref{sec:1} we derive the predictive models for the drag coefficient of ellipsoidal particles with different aspect ratios immersed in a uniform ambient flow. Such models include rarefaction, orientation, and gas-surface interaction effects, and we show they correctly reproduce DSMC data on a broad range of $Kn$. In Section \ref{sec:2}, we modify the simulation setup adding solid walls to the simulation box boundaries perpendicular to the vertical direction, and we locate the ellipsoidal particle in proximity of the top wall. We show that the presence of the wall increases the effective drag experienced by the particles, leading to a deviation from the sine-squared drag law typical of the unbounded case. Since such effects quickly vanish at increasing Knudsen and particle-wall distance, drag corrections from Section \ref{sec:1} can be applied also in confined flow domains conserving good accuracy (assuming large particle-based Knudsen and that the size of the particle is much smaller than the typical system size). Finally, in Section \ref{sec:conc}, we summarize and discuss our results.

\section{Effects of the aspect ratio and gas-surface interactions at finite Knudsen number}
\label{sec:1}
In our previous work \cite{livi} we have shown that it is possible to model orientation and rarefaction effects on simple ellipsoidal particles immersed in a uniform ambient flow by using a perturbative approach where rarefaction effects are modeled as a continuous function of Knudsen. Such function is obtained through a fit of the drag force experienced by the particles, as measured from DSMC simulations, for the two cases with orientation at $\Phi=0^\circ$ and $\Phi=90^\circ$, with respect to the uniform ambient flow. Orientation effects are then extrapolated for all values of $\Phi$ through the sine-squared drag law [Eq. (\ref{eq:sine_squared_f})], which we show is still valid also for rarefied conditions. Since the discussion presented in \cite{livi} will be the starting point of this work, we will briefly summarize the main concepts. We proposed a model for the drag coefficient on ellipsoidal particles based on the following expression:
\begin{align}
C_{D,\chi}(Kn) = \underbrace{C_{D,\chi}^{cont}}_{\mbox{continuum}} \cdot \underbrace{g_{\chi}(Kn)}_{\mbox{rarefaction effects}},
\label{eq:ellipsoid_prediction}
\end{align} 
where the subscript $\chi$ refers to the two cases  with $\Phi=0^\circ$ and $\Phi=90^\circ$, separately, $C_{D,\chi}^{cont}$ is the drag coefficient of the ellipsoidal particles under investigation in the continuum regime and $g_{\chi}(Kn)$ is the model function to be determined.\\
\begin{figure}[h!]
\centering
\begin{tikzpicture}

\node[anchor=south west,inner sep=0] (image) at (0,0) {\includegraphics[width=0.45\textwidth]{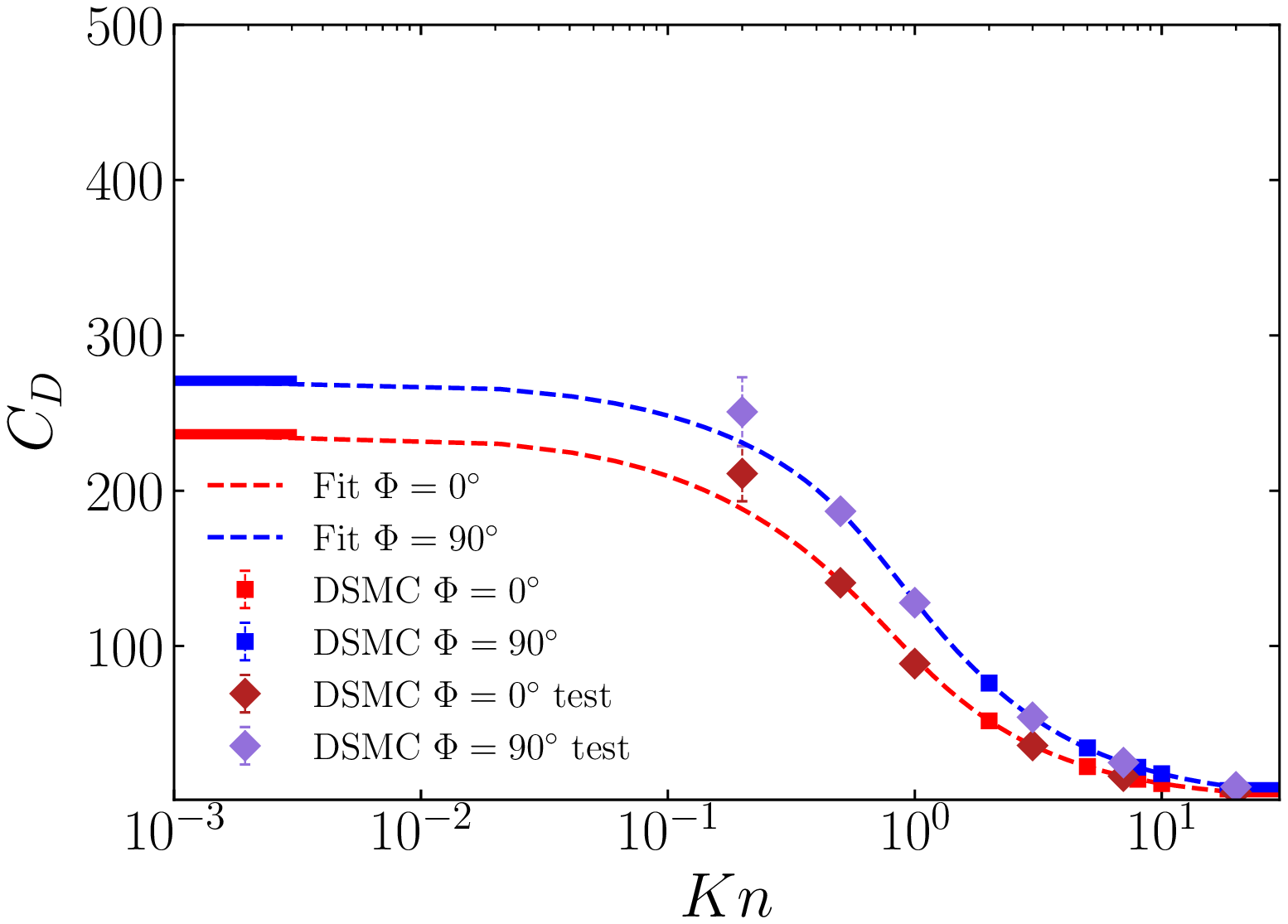}};
\node[anchor=south west,inner sep=0] (image) at (4.90,3.45) {\includegraphics[width=0.175\textwidth]{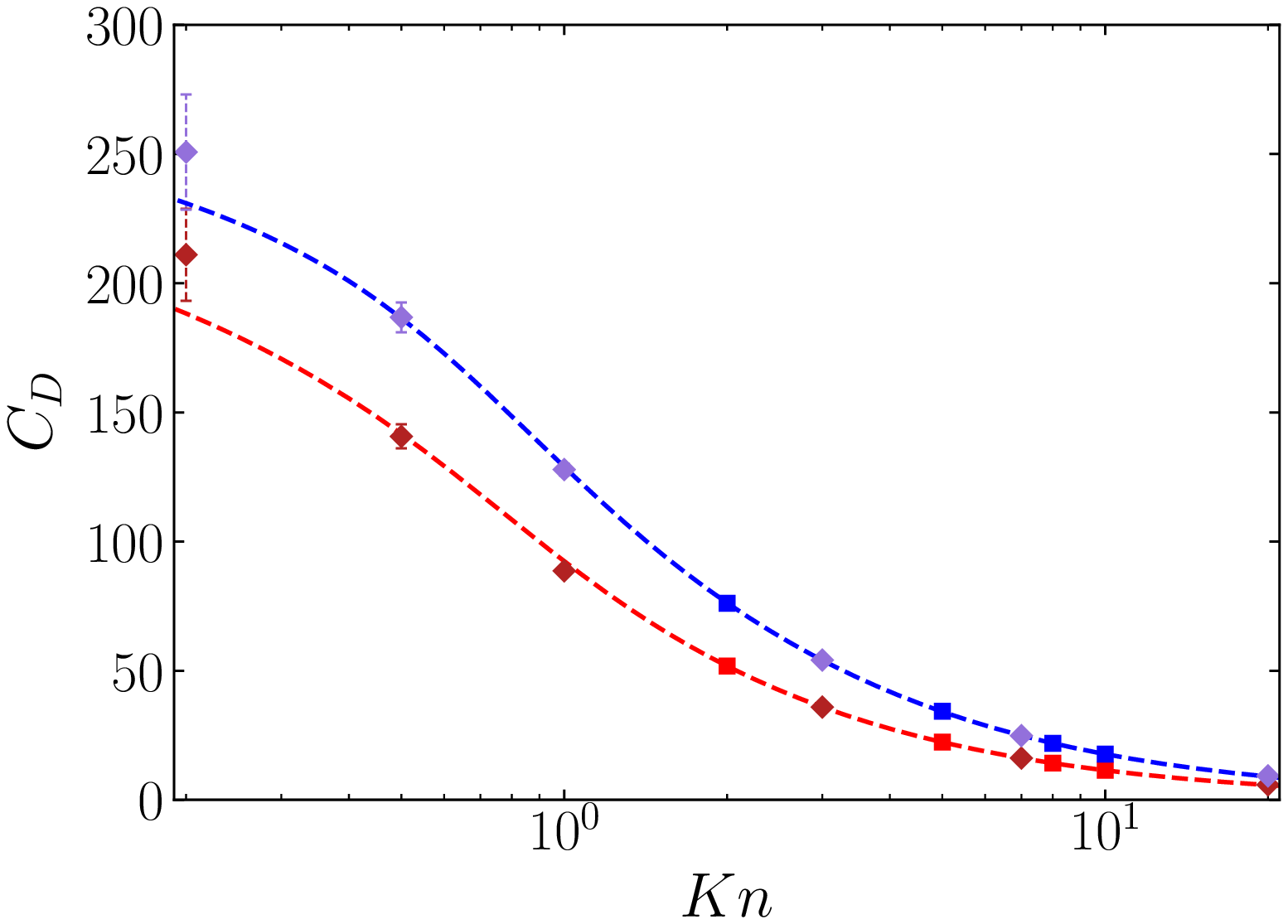}};
\end{tikzpicture}\\
\iffalse
\begin{tikzpicture}

\node[anchor=south west,inner sep=0] (image) at (0,0) {\includegraphics[width=0.45\textwidth]{newmodel_fit_oblate_2.eps}};
\node[anchor=south west,inner sep=0] (image) at (4.85,3.45) {\includegraphics[width=0.175\textwidth]{newmodel_fit_oblate_2_zoom.eps}};
\end{tikzpicture}\\

\caption{\small{Fit of DSMC simulation data of $ C_{D,0^\circ}$ (red) and $C_{D,90^\circ}$ (blue) using the functions in Eqs. (\ref{eq:ellipsoid_prediction})and (\ref{eq:g}), for a prolate ellipsoid (top) and an oblate ellipsoids (bottom) with aspect ratio $a/b=2$ for fully diffusive reflections ($\sigma=1$). The  predictive models (dashed lines) correctly captures data from DSMC simulations that is not used during the fitting process (colored diamonds), as well as the continuum and free molecular limits (horizontal solid lines). The free molecular limit is calculated using the relation from \cite{dahneke3}.}}
\fi
\caption{\small{Fit of DSMC simulation data (squares) of $ C_{D,0^\circ}$ (red) and $C_{D,90^\circ}$ (blue) using the functions in Eqs. (\ref{eq:ellipsoid_prediction}) and (\ref{eq:g}), for a prolate ellipsoid with aspect ratio $a/b=2$ and a fully diffusive surface ($\sigma=1$). The  predictive models (dashed lines) correctly captures data from $Kn_{test}$ set (diamonds) down to $Kn=0.5$, while they start to deviate at $Kn=0.2$ (approaching slip-flow regime). The continuum and free molecular limits (horizontal solid lines) are correctly reproduced by the models. The former is obtained from the spherical case using the relations derived by Oberbeck \cite{oberbeck}, while the latter is calculated from \cite{dahneke3}. The inset plot shows a zoom on the DSMC data and asymptotic limits are omitted. Error bars are calculated from $\varepsilon_{95}$.}}
\label{fig:cont_recovery}
\end{figure}
\begin{figure*}
\centering
\includegraphics[width=0.32\textwidth]{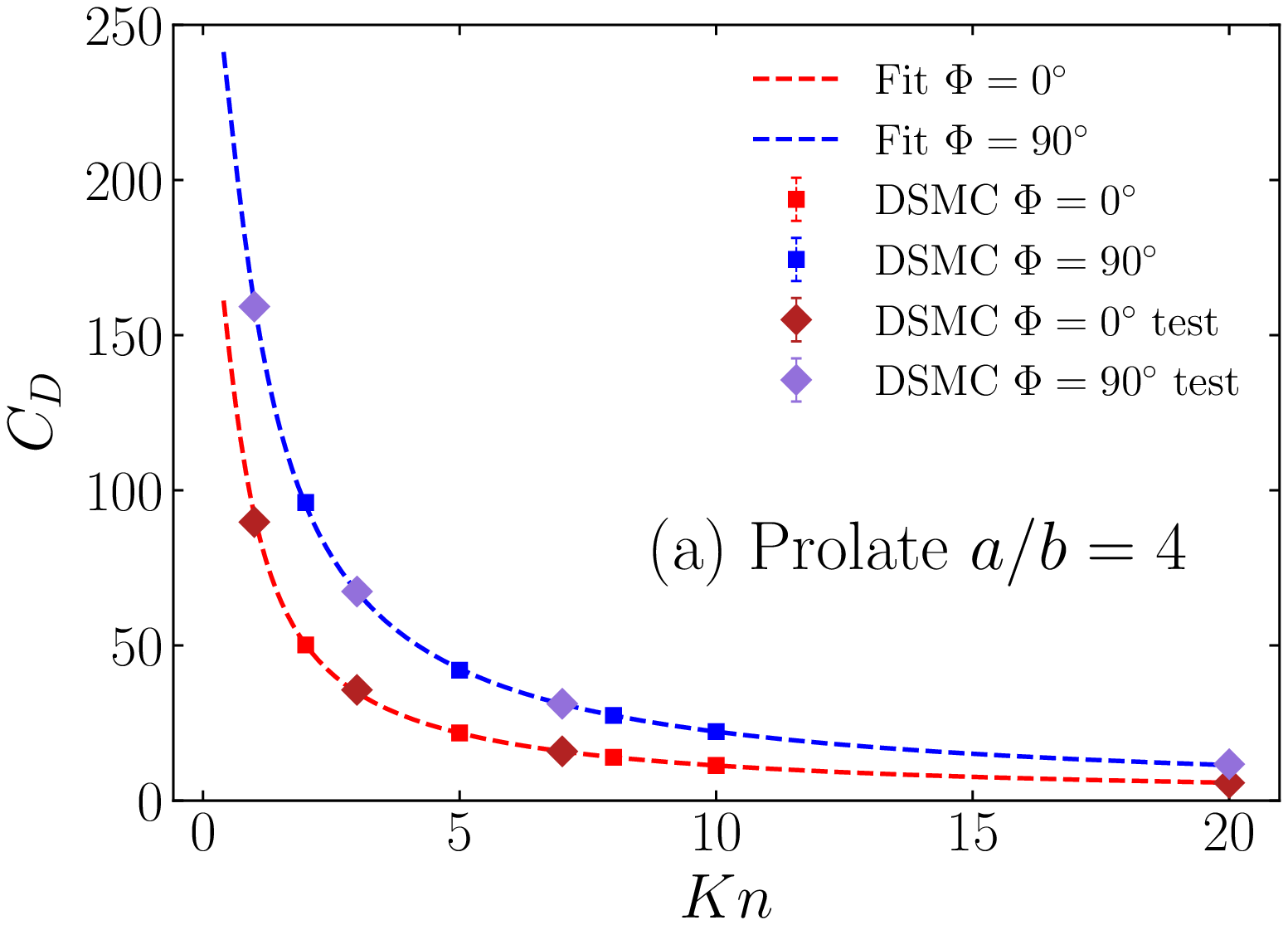}%
\includegraphics[width=0.32\textwidth]{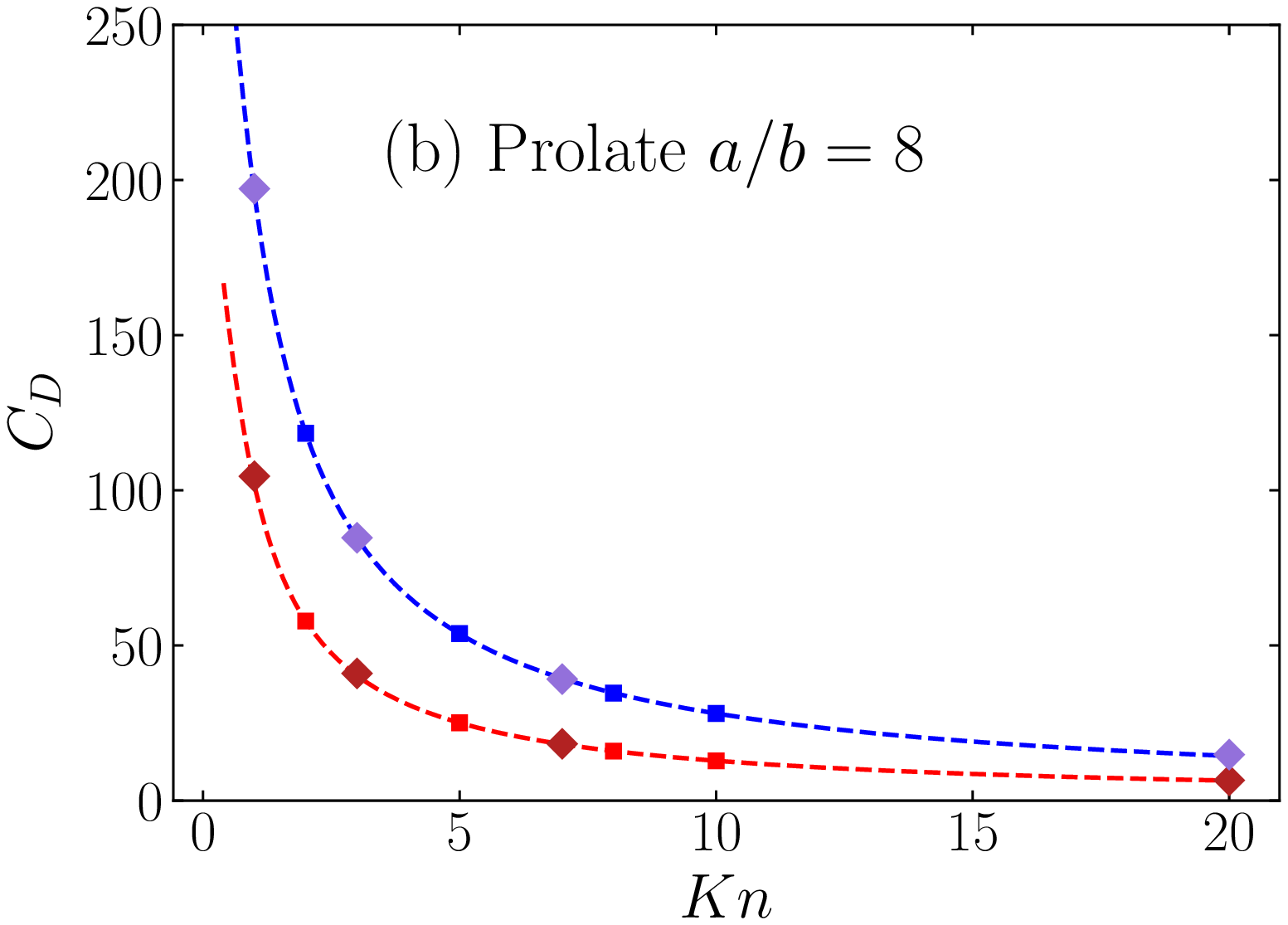}%
\includegraphics[width=0.32\textwidth]{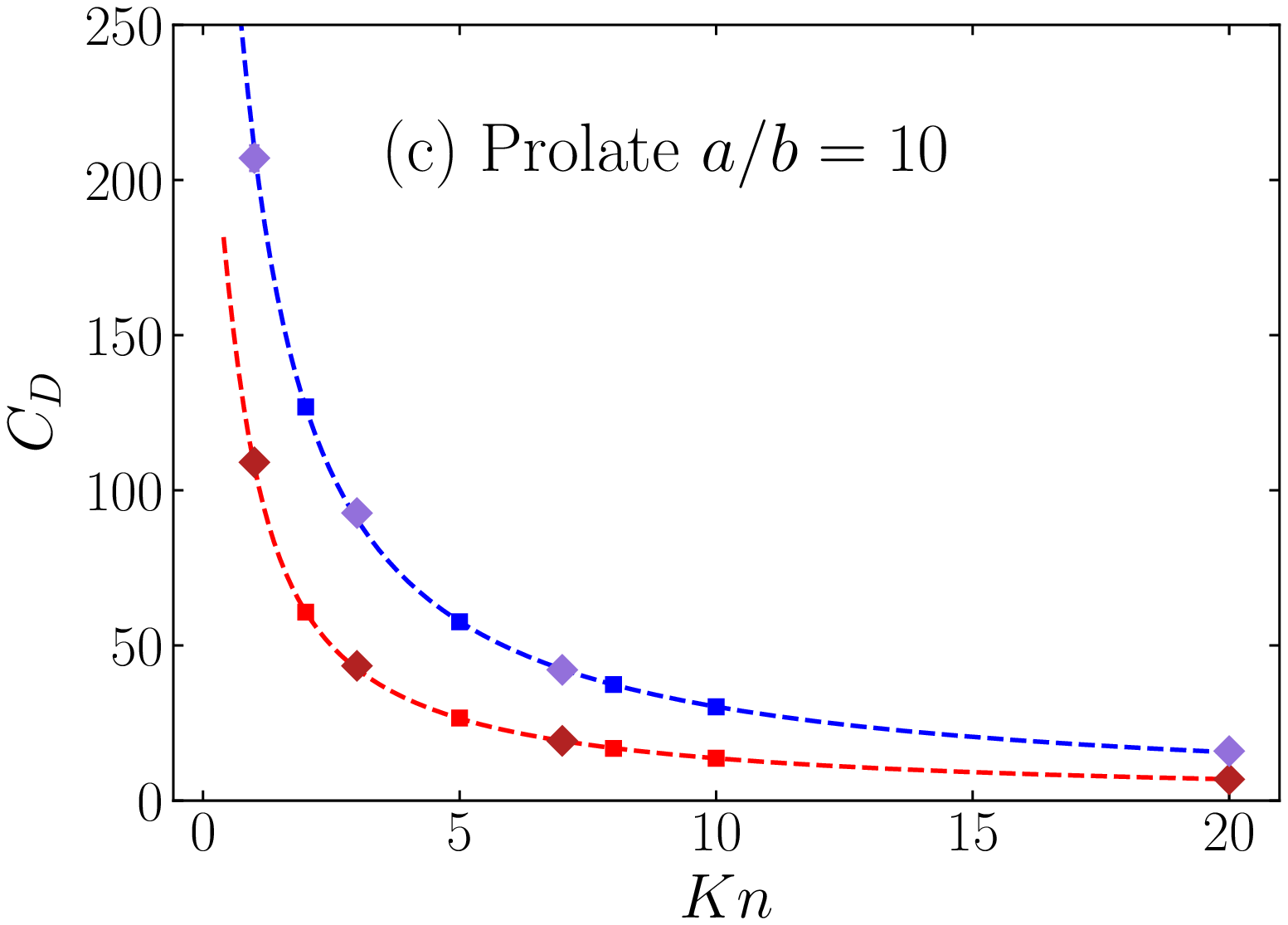}\\
\includegraphics[width=0.32\textwidth]{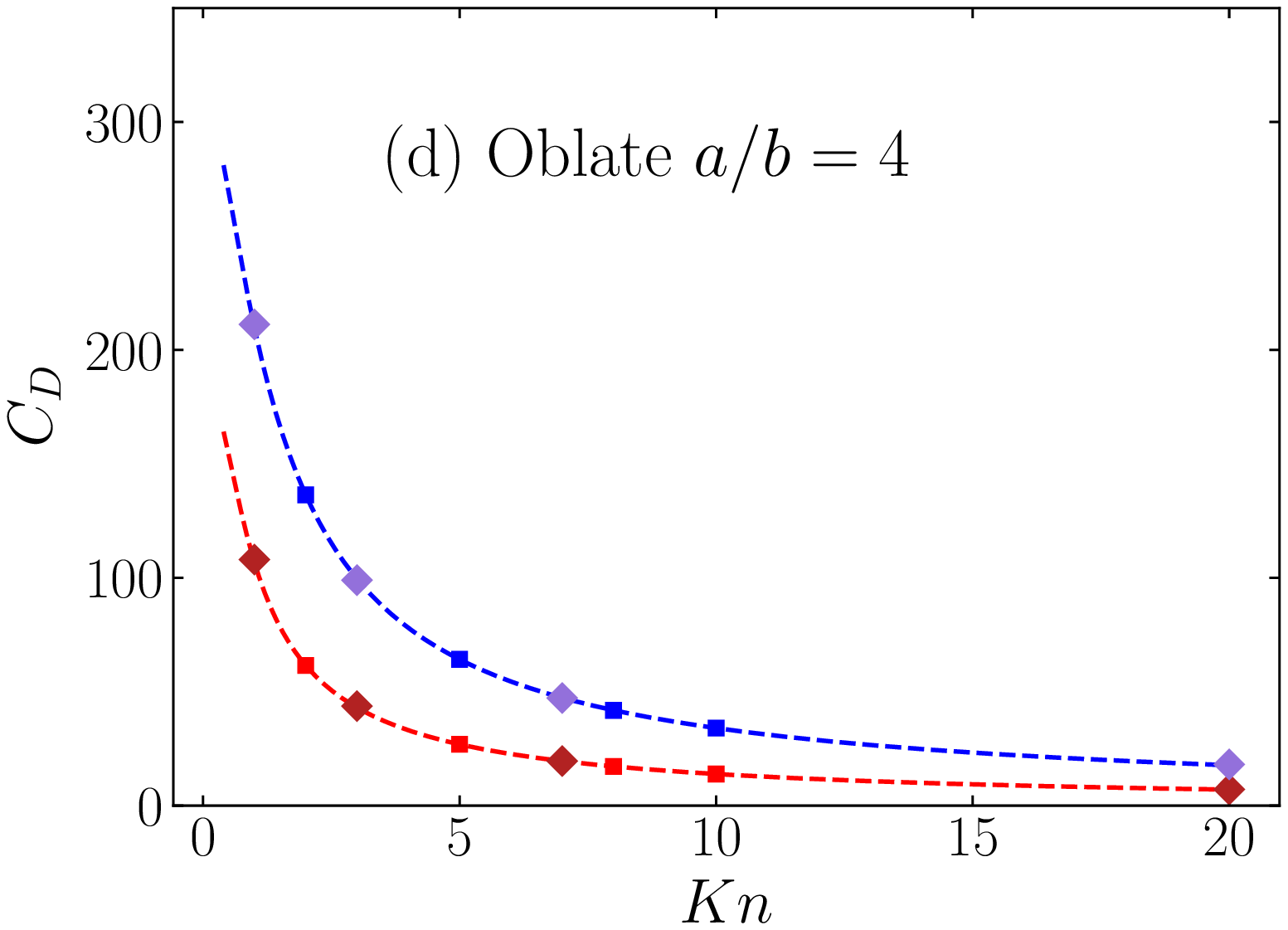}%
\includegraphics[width=0.32\textwidth]{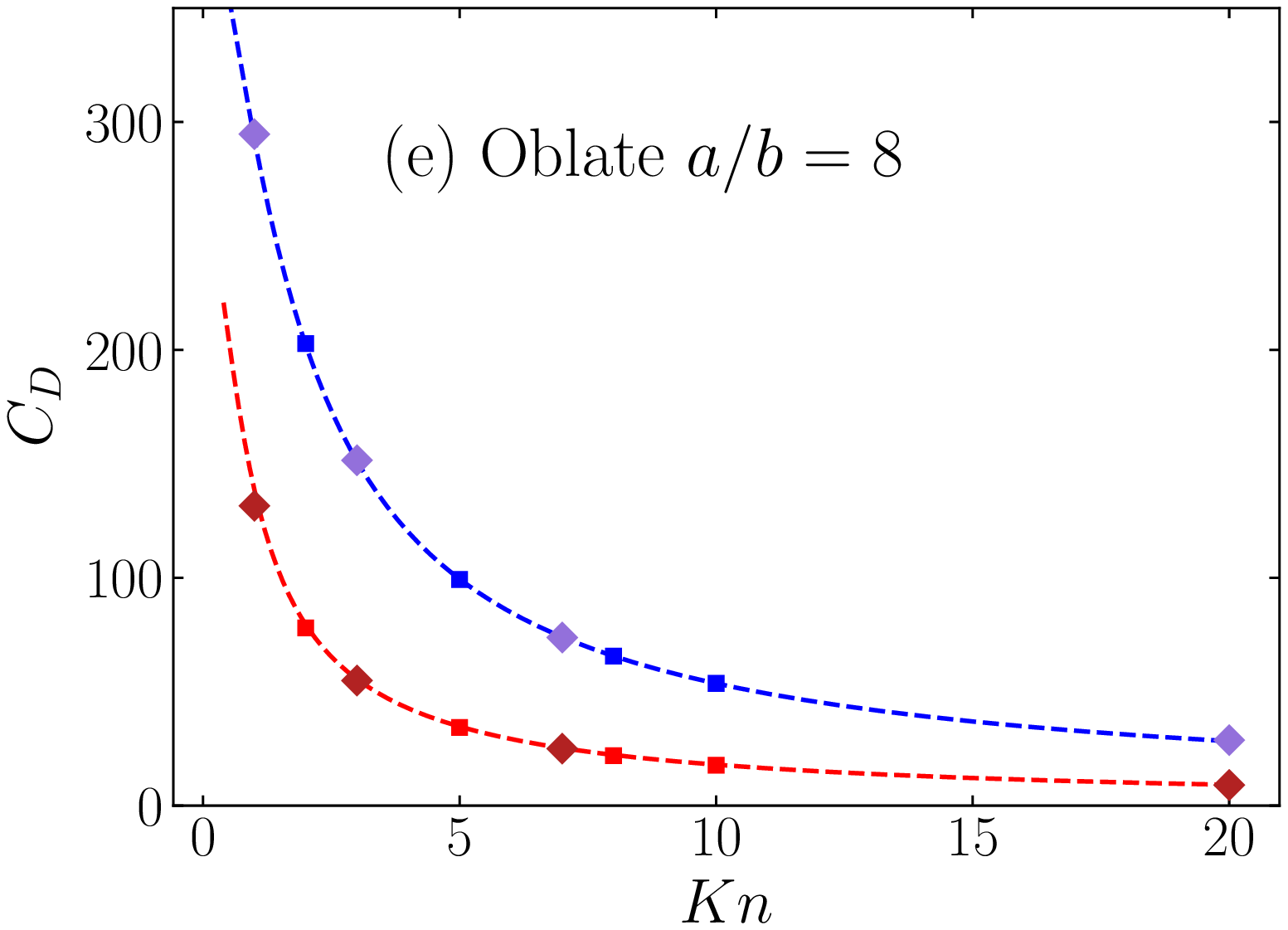}%
\includegraphics[width=0.32\textwidth]{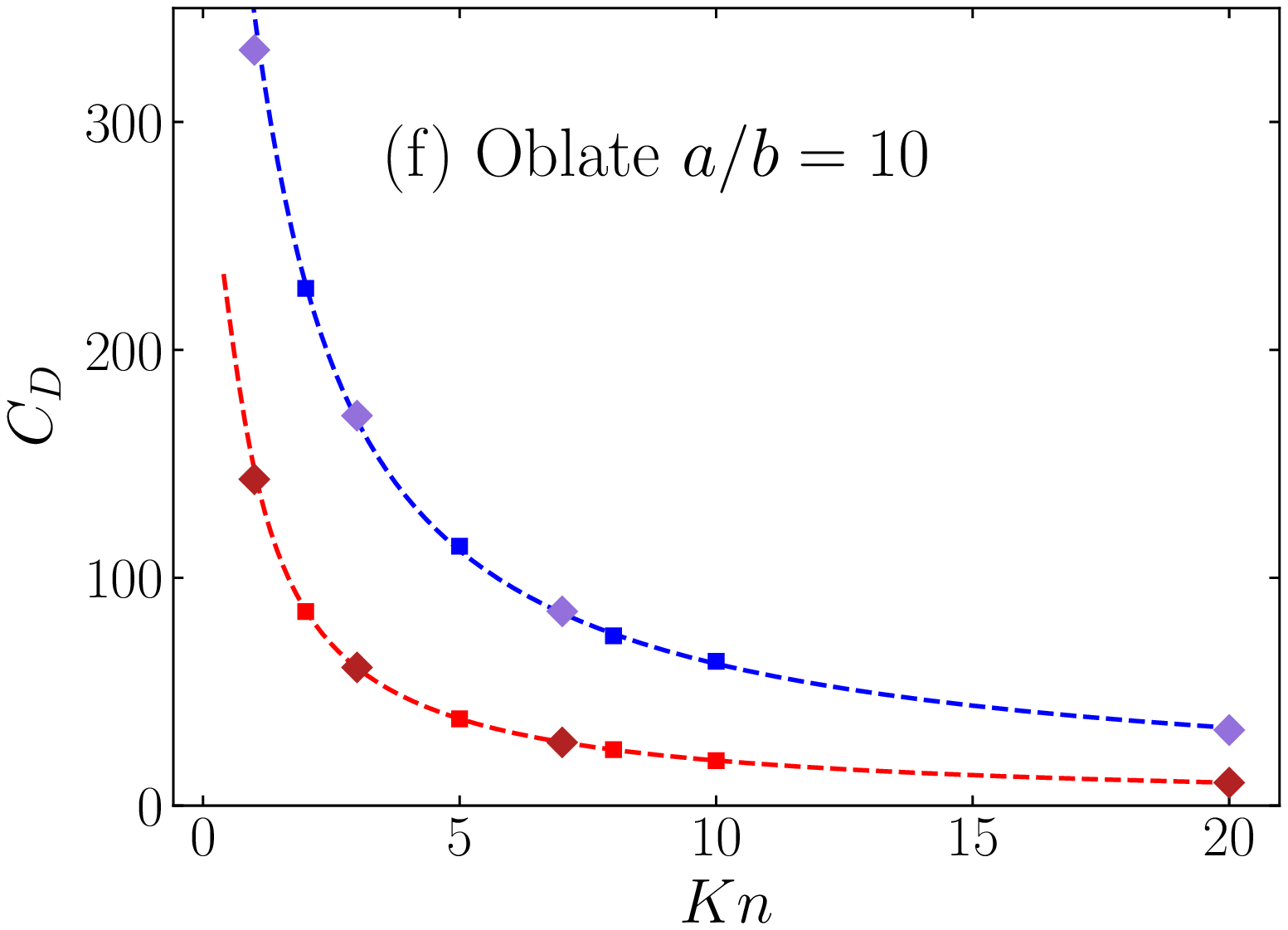}%
\caption{\small{Fit of DSMC simulation data of $ C_{D,0^\circ}$ (red) and $C_{D,90^\circ}$ (blue) using Eqs. (\ref{eq:ellipsoid_prediction}) and (\ref{eq:g}), for prolate ellipsoids (top row) and oblate ellipsoids (bottom row) with same volume and $a/b=4$ (a,d), $a/b=8$ (b,e) and $a/b=10$ (c,f). Fitted curves (dashed lines) represent the model functions given by Eq. (\ref{eq:g}) and using Eq. (\ref{eq:ellipsoid_prediction}). The prediction from the derived models is compared with data from DSMC simulations that is not used during the fitting process (colored diamonds), showing an excellent match. As shown in \cite{livi}, it is sufficient to know the scaling of the drag force at $\Phi=0^\circ$ and $\Phi=90^\circ$ to derive its value at any arbitrary orientation. Legend is shown only for (a) as symbols refer to same quantities in all plots. }}
\label{fig:drag_ab}
\end{figure*}
 \begin{figure}[h!]
\centering
\begin{tikzpicture}

\node[anchor=south west,inner sep=0] (image) at (0,0) {\includegraphics[width=0.45\textwidth]{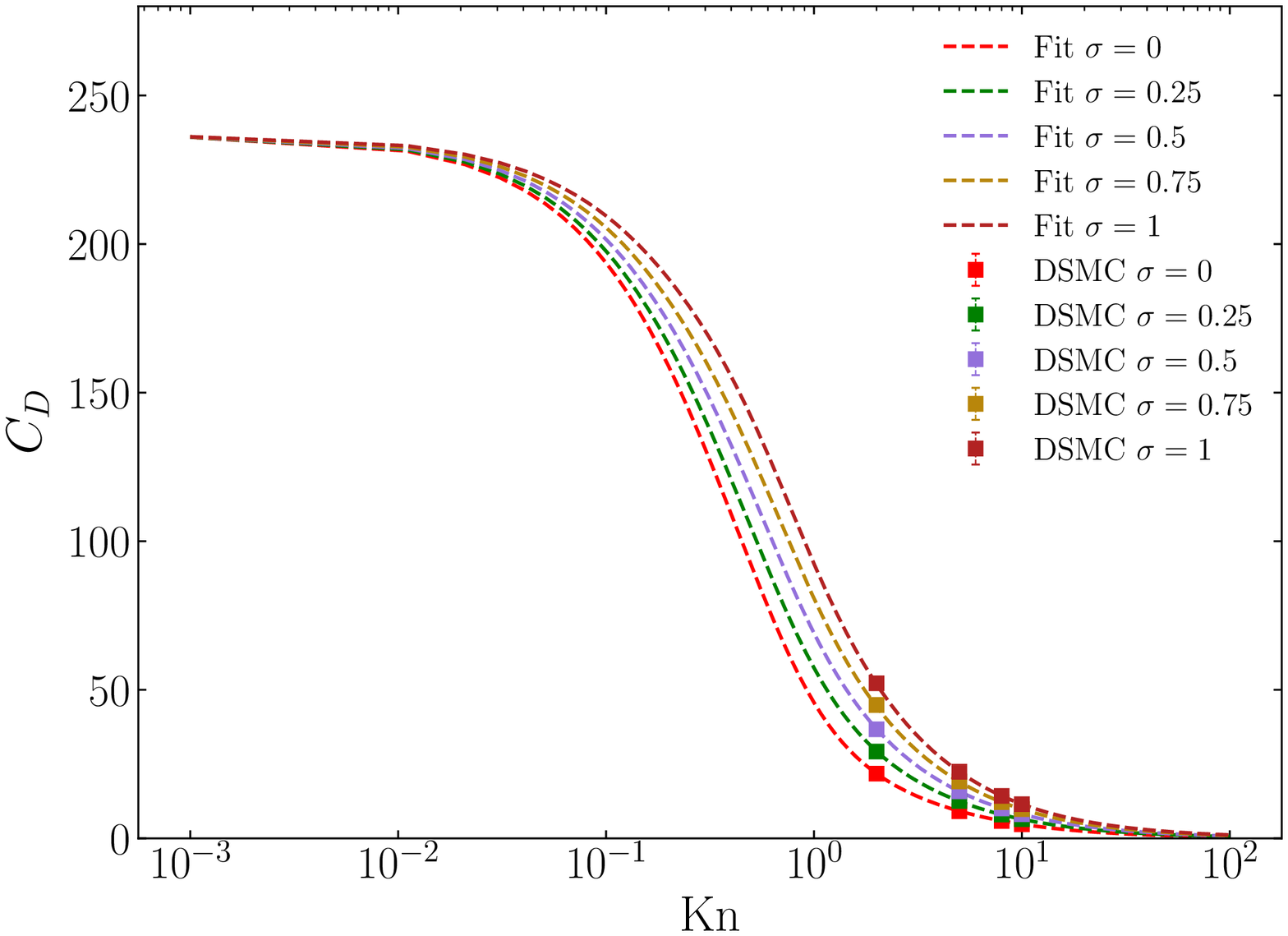}};
\node[anchor=south west,inner sep=0] (image) at (0.98,0.93) {\includegraphics[width=0.185\textwidth]{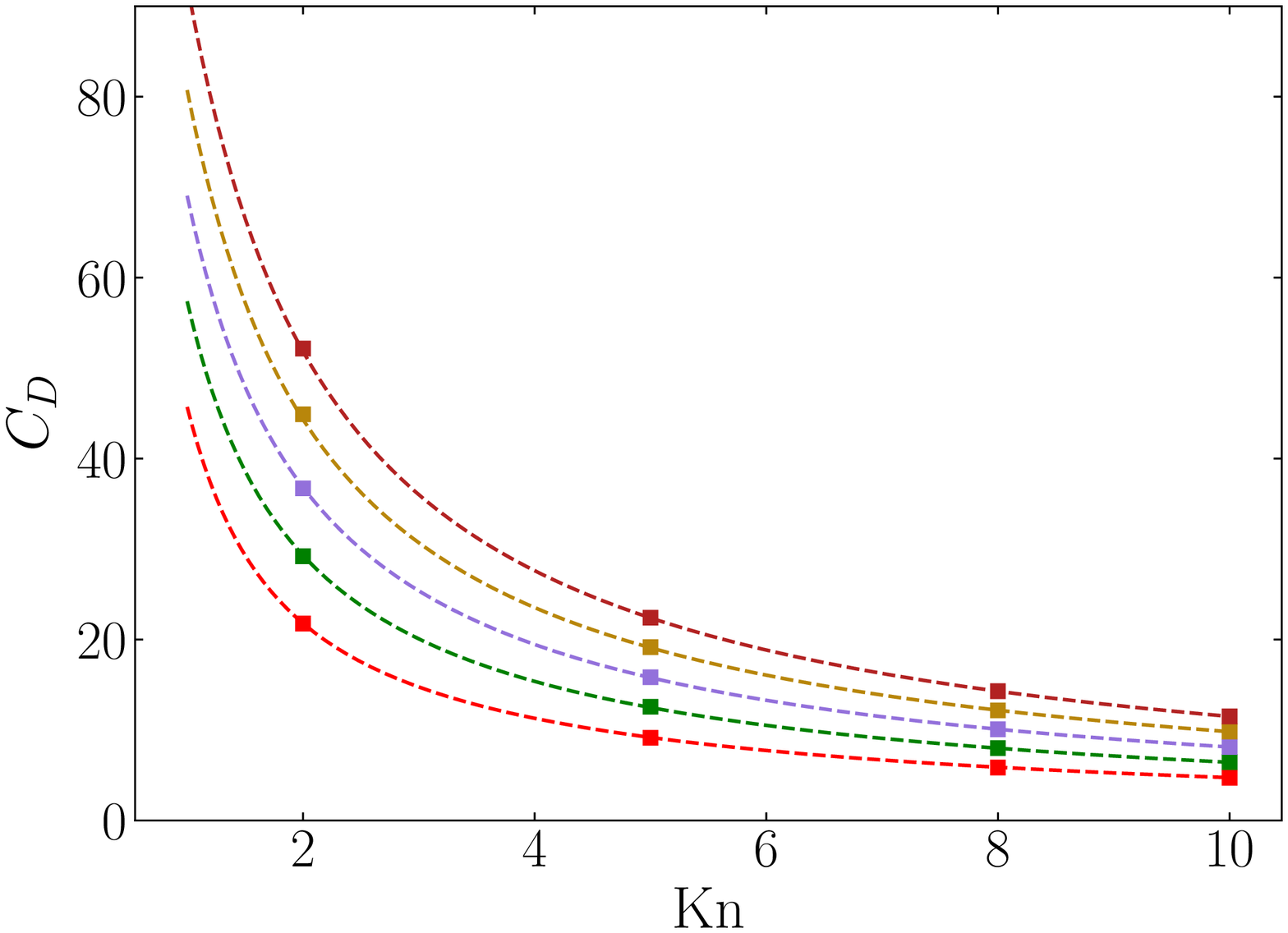}};
\end{tikzpicture}\\

\begin{tikzpicture}

\node[anchor=south west,inner sep=0] (image) at (0,0) {\includegraphics[width=0.45\textwidth]{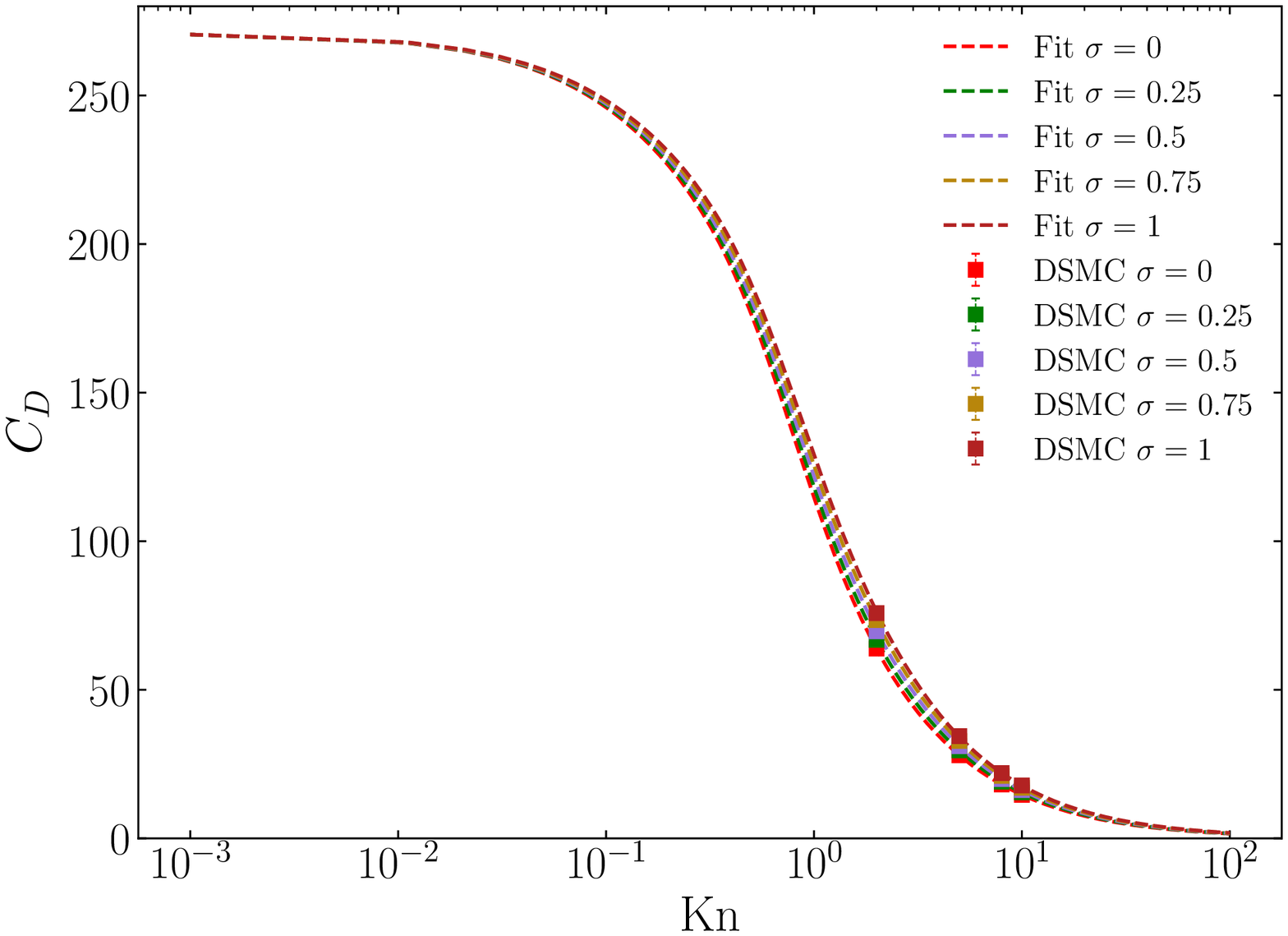}};
\node[anchor=south west,inner sep=0] (image) at (0.98,0.93) {\includegraphics[width=0.185\textwidth]{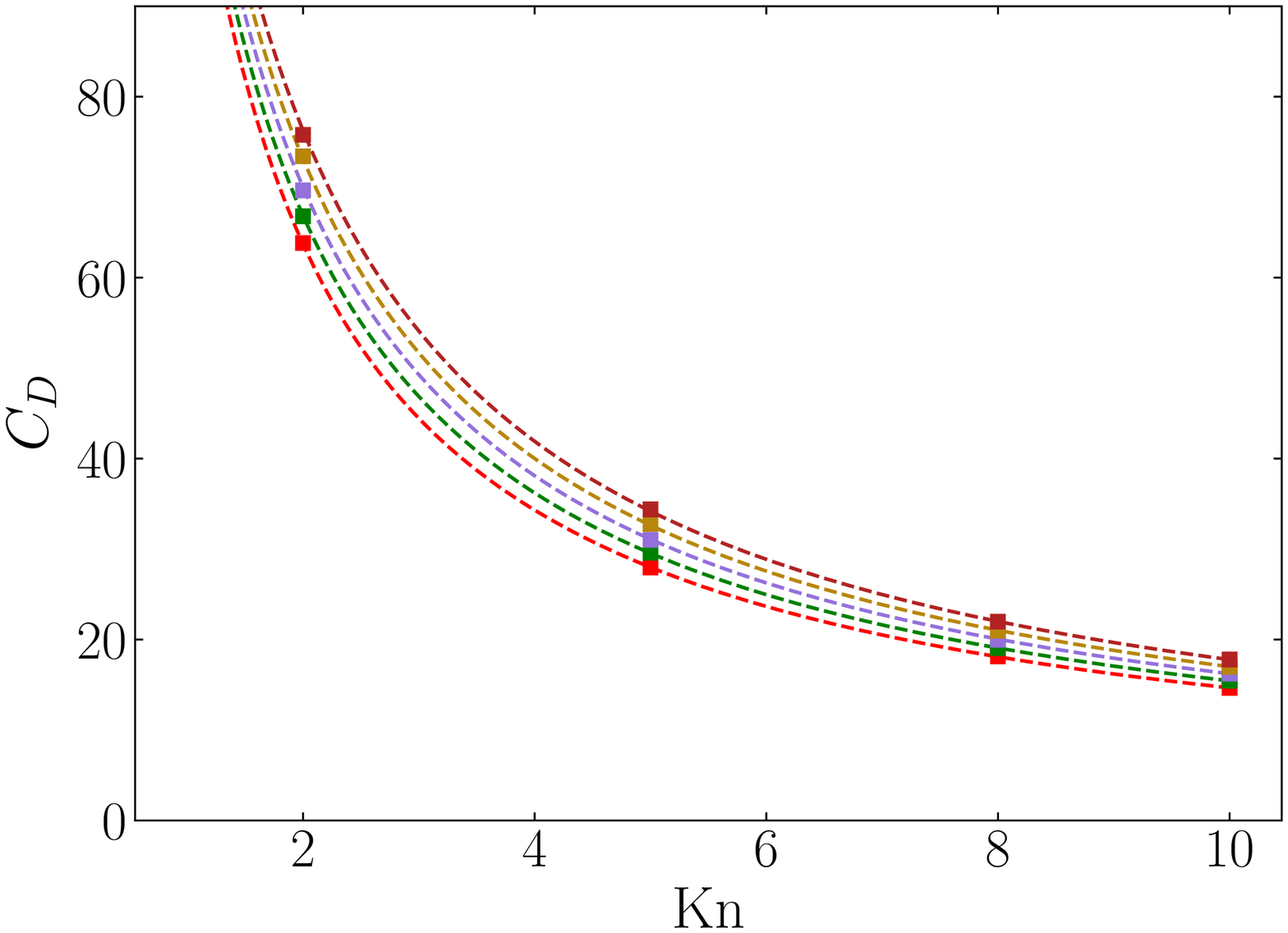}};
\end{tikzpicture}\\

%\caption{\small{Derivation of the model functions from Eqs. (\ref{eq:g_0_tmac})-(\ref{eq:g_90_tmac}) (dashed lines) for a prolate (top) and oblate (bottom) ellipsoidal particle with aspect ratio $a/b=2$. Model functions are derived through the fit of DSMC data (squares) for different values of $sigma$ and $Kn$. }}

\caption{\small{Fit of DSMC simulation data (squares) of $ C_{D,0^\circ}$ (top) and $C_{D,90^\circ}$ (bottom) using Eqs. (\ref{eq:ellipsoid_prediction}) and (\ref{eq:g_tmac}) (dashed lines) for a prolate ellipsoidal particle with aspect ratio $a/b=2$. The model functions are derived through the fit of DSMC data using  $\sigma_{fit}$ and $Kn_{fit}$. In the insets, an enlargement on the $Kn_{fit}$ range is presented. }}
\label{fig:drag_tmac}
\end{figure}
The drag coefficient in Eq. (\ref{eq:ellipsoid_prediction}) is computed from the spherical case using the relations derived by Oberbeck \cite{oberbeck}, while the particle-based Knudsen number is defined as $Kn = \lambda/R_{eq}$, where $\lambda$ is the mean free path of the gas, and $R_{eq}$ is the radius of the sphere with equivalent volume. In \cite{livi} we showed that this definition of $Kn$ captures rarefaction effects on ellipsoidal particles unambiguously, and we proposed the following form for the model functions:
\begin{align}
g_{\chi}(Kn) = f(Kn) + \frac{a_{\chi}}{b_{\chi} + c_{\chi}Kn},
\label{eq:g_old}
\end{align}
where $a_\chi,\ b_\chi$ and $c_\chi$ are free parameters determined through a fit of data from DSMC simulations, and $f(Kn)$ represents the rarefaction correction to the Stokes drag force for a spherical particle as proposed by Phillips \cite{phillips} and given by
\begin{equation}
f(Kn) = \frac{15 - 3c_1 Kn + c_2(8+\pi \sigma)(c_1^2 +2)Kn^2}{15 +12c_1 Kn + 9 (c_1^2 +1)Kn^2 + 18c_2(c_1^2+2)Kn^3}, 
\label{eq:f_phil}
\end{equation} 
with $c_1 = \frac{2-\sigma}{\sigma}$, $c_2 = \frac{1}{2-\sigma}$. It can be seen that Eq. (\ref{eq:f_phil}) recovers the continuum limit for vanishing $Kn$ as
\begin{align}
\lim_{Kn \to 0 }f(Kn)= 1.
\label{eq:f_lim}
\end{align}
While this model is able to predict with good accuracy the effects of rarefaction in the transition and free-molecular regime of ellipsoidal particles, it is limited to a single aspect ratio ($a/b=2$), to fully diffuse reflection ($\sigma=1$), and it does not properly recover the asymptotic limit for the continuum regime.\\
In this Section we propose an improved model to address the aforementioned aspects, starting from the recovery of the correct asymptotic continuum limit. The new functional form for the model functions is given by:
\begin{align}
g_{\chi}(Kn) = f(Kn) + \frac{p_{\chi}Kn}{q_{\chi} + r_{\chi}Kn^{s_{\chi}}},
\label{eq:g}
\end{align}
where $p_\chi,\ q_\chi,\ r_\chi$ and $s_\chi$, with $\chi=0^\circ$ or $90^\circ$, are the new free parameters to be determined separately for $C_{D,0^\circ}$ and $C_{D,90^\circ}$ and for each different shape of the particle. The free parameter $s_\chi$ is set to be strictly larger than unity, so that
\begin{align}
\lim_{Kn \to 0 }g_{\chi}= 1,
\label{eq:g_lim}
\end{align}
\begin{align}
\lim_{Kn \to +\infty}g_{\chi}= 0,
\label{eq:g_lim}
\end{align}
 allowing to correctly recover the continuum and free molecular limits also for the ellipsoidal case.\\
To obtain the free parameters in Eq. (\ref{eq:g}), we perform collisional DSMC simulations of different ellipsoidal particles immersed in uniform argon gas flow with ambient velocity $U_0$, varying $Kn$ and orientation $\Phi$, following a similar procedure as described in \cite{livi} (a sketch of the simulation setup is presented in Fig. \ref{fig:sk_1}). The volume of the equivalent sphere is fixed to $V=6.5\times 10^{-20}\mbox{m}^3$, corresponding to $R_{eq}=2.5\mu\mbox{m}$. The aspect ratio of the particles is initially set to $a/b=2$, leading to a major radius $a = 0.39 \mu \mbox{m}$ for the prolate case and $a = 0.315 \mu \mbox{m}$ for the oblate case (later in the paper the aspect ratio will be varied to $a/b=4,8$ and $10$). The physical simulation box size is set to $L=20a $. In terms of DSMC cell units, for all cases with $Kn\geq 2$ a grid resolution of $120$ cells per linear direction is sufficient to ensure high accuracy accordingly to the DSMC rules of thumb  \cite{bird,garcia} (i.e. $C_{size}\leq 0.3 \lambda$, where $C_{size}$ is the size of a single DSMC grid cell). For simulations at $Kn=1,0.5$ and $0.2$ the number of cells is increased to $144,288$ and $640$, respectively. The number of computational particles-per-cell (PPC) is always set to be larger than $25$, again in accordance with \cite{bird,garcia}. The Reynolds number is fixed to $Re=0.1$, so that the ambient flow, $U_0$, the gas density, $\rho$, and the pressure are obtained from the values of $Kn$ and $Re=\rho2U_0R/\mu$, where $\mu =2.12\times 10^{-5}\mbox{ kg m}^{-1} \mbox{s}^{-1}$ is the dynamic viscosity of the gas. The temperature is set to $T=300K$. The TMAC is initially kept constant with $\sigma = 1$ (fully diffuse reflections) and later it will be varied in the whole range $0\leq \sigma \leq 1$. The error bars on the measured drag are calculated using the $95\%$ confidence interval defined as $\varepsilon_{95} = 2\sigma_{std}/\sqrt{N_{\Delta t}}$, where $\sigma_{std}$ is the standard deviation on the average value of the drag force from DSMC simulations, $F_D$, and $N_{\Delta t}$ is the number of samples, which is set to $N_{\Delta t}=10000$ for $Kn\geq 0.5$, $N_{\Delta t}=50000$ otherwise. Sampling of the drag force start once the steady state is reached. The available DSMC data is divided in two sets: $Kn_{fit} = 2,5,8,10$, which is used as data points for the fit, and $Kn_{test}=0.2,0.5,1,3,7,9,20$ which is instead used to validate the model. \\
We derive the fit parameters using the same procedure discussed in \cite{livi} for the cases of a prolate and oblate ellipsoid with $a/b=2$, showing that the new model correctly predicts $C_D$ while recovering the asymptotic continuum limit. Results, limited to the prolate case, are presented in Fig. \ref{fig:cont_recovery}.\\
It is important to mention that while this derivation shows robust performances in predicting rarefaction corrections to the drag experienced by ellipsoidal particles down to values of $Kn=0.5$, it is not expected to be accurate for the slip-flow regime ($Kn\lesssim 0.1$) as the $Kn_{fit}$ set is limited to the transition and free-molecular regimes. The loss of accuracy of the model for low values of Knudsen can be observed in Fig. \ref{fig:cont_recovery}, where the model predictions start to deviate from DSMC test data at $Kn=0.2$. In order to improve the model in the slip-flow regime, more fitting data should be provided, either numerically or experimentally and currently our model should be used only for $Kn>0.2$. \\
\begin{figure*}
\centering
\includegraphics[width=0.2\textwidth]{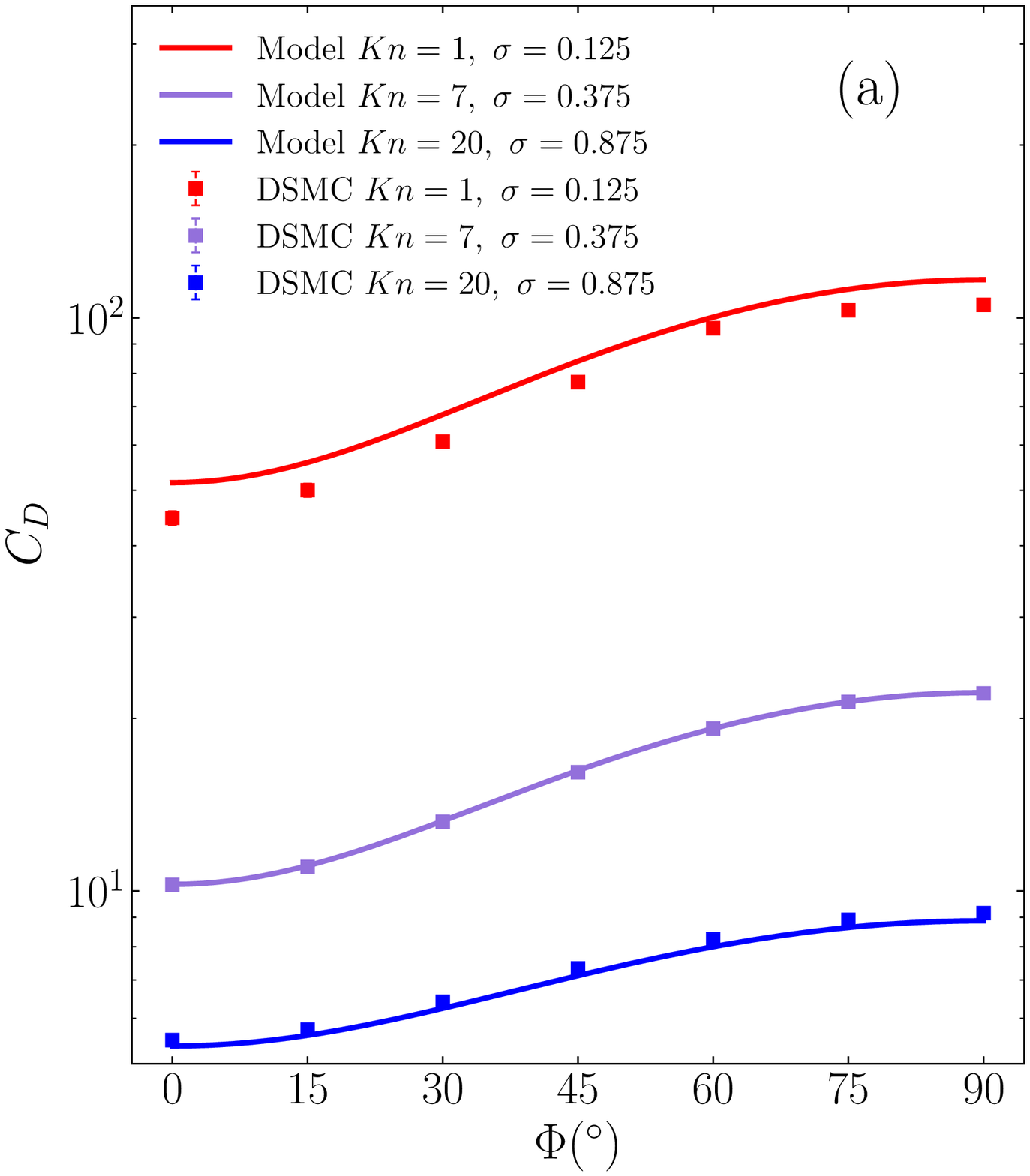}
\includegraphics[width=0.2\textwidth]{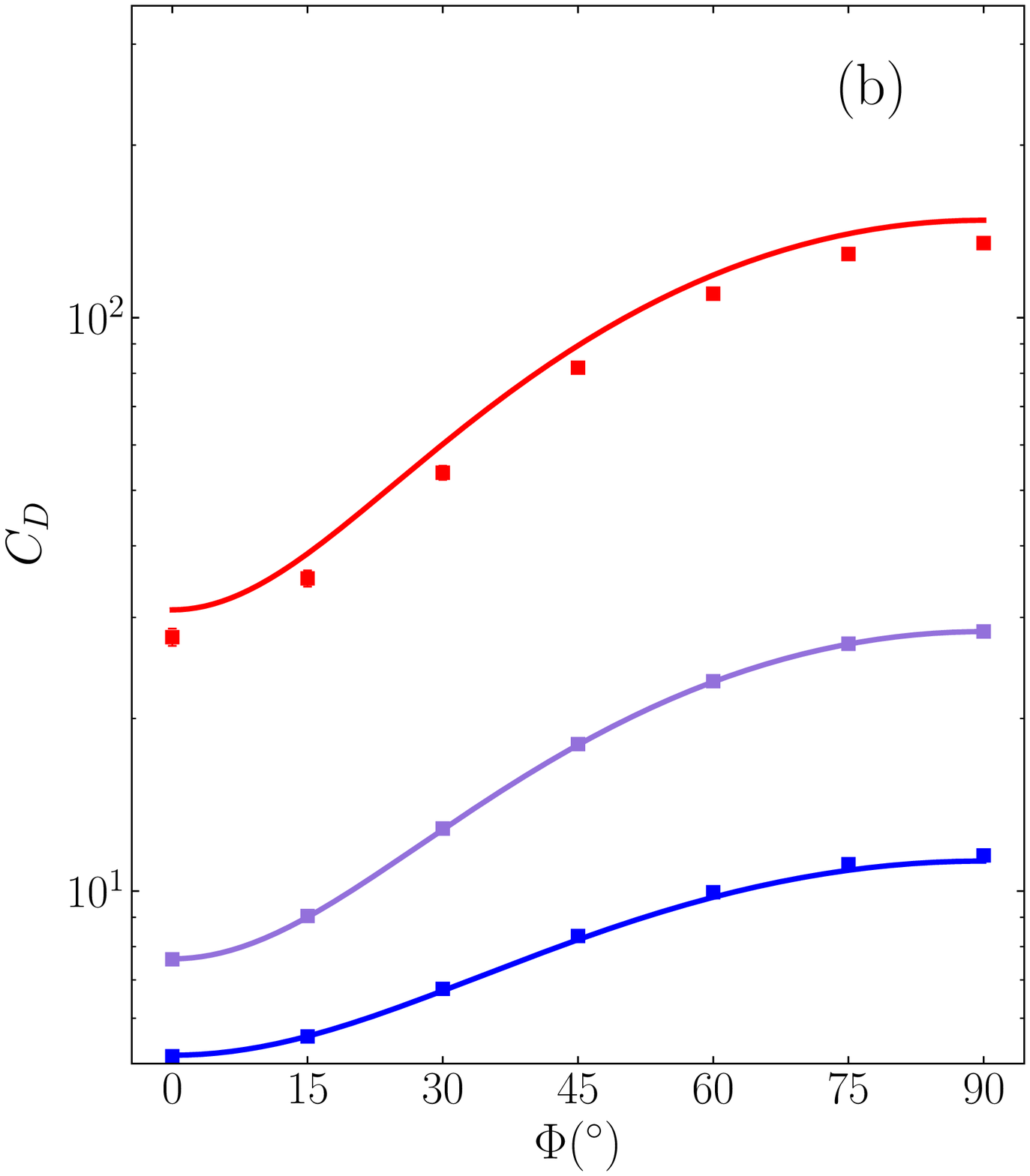}
\includegraphics[width=0.2\textwidth]{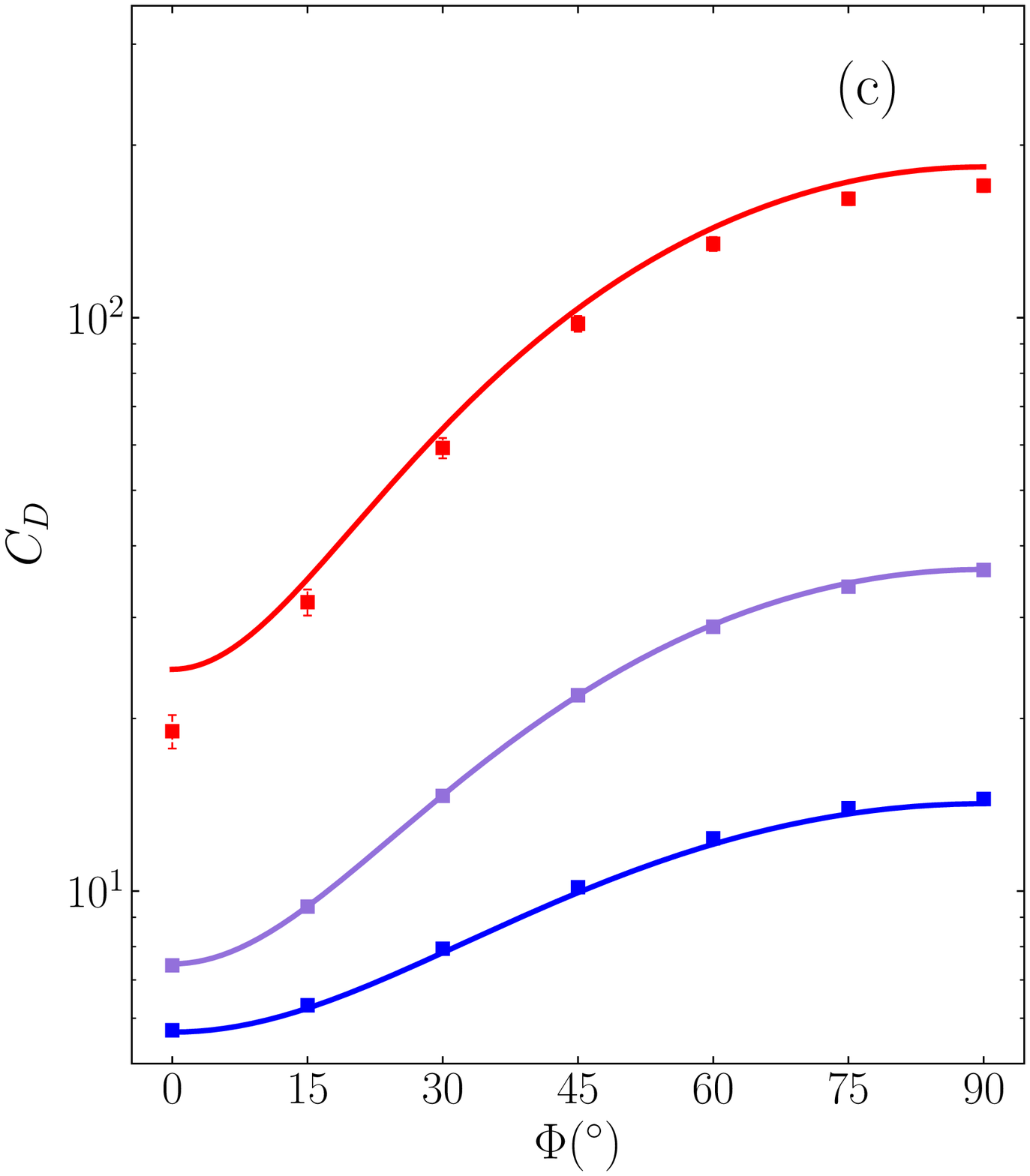}
\includegraphics[width=0.2\textwidth]{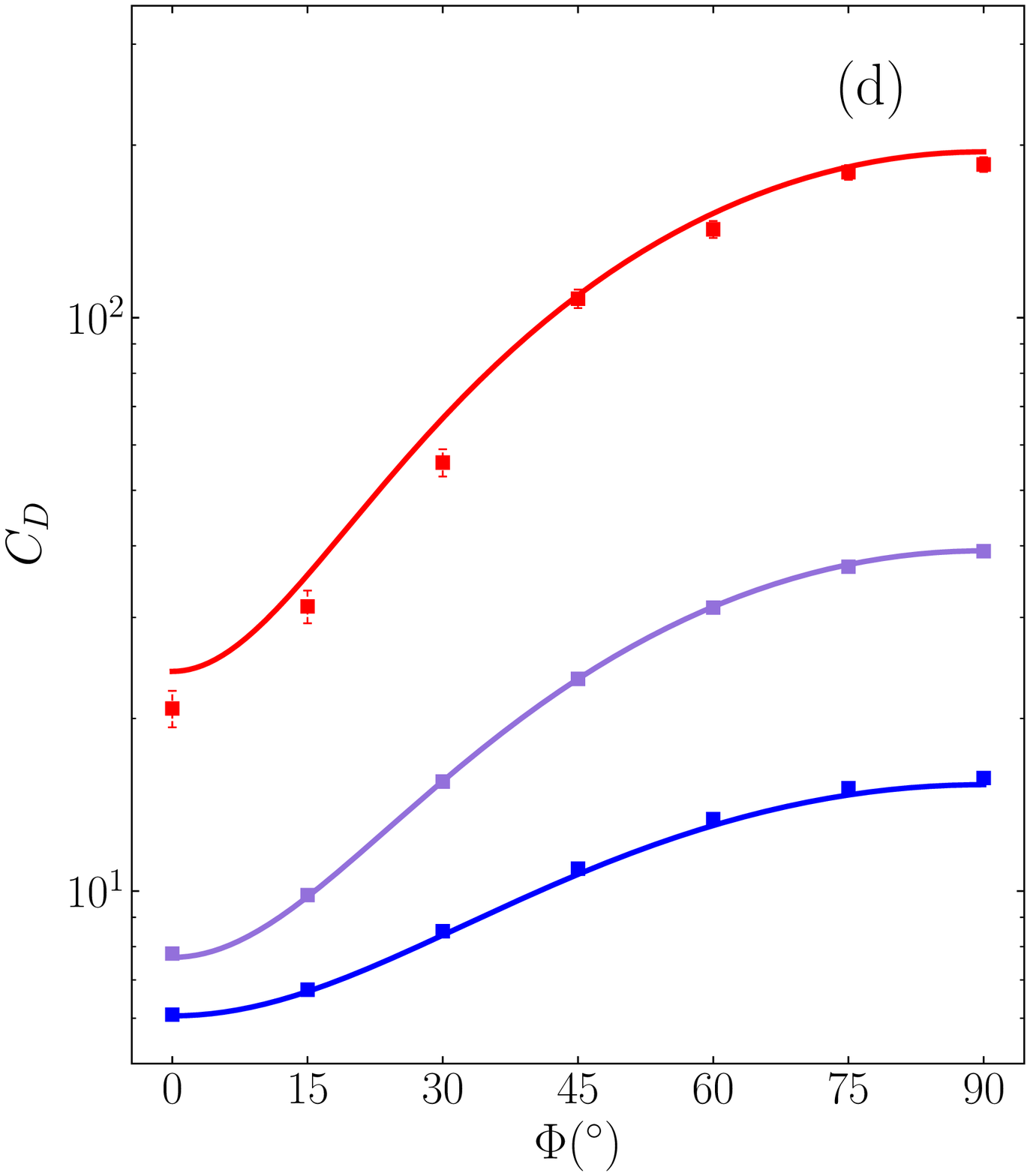}\\
\includegraphics[width=0.2\textwidth]{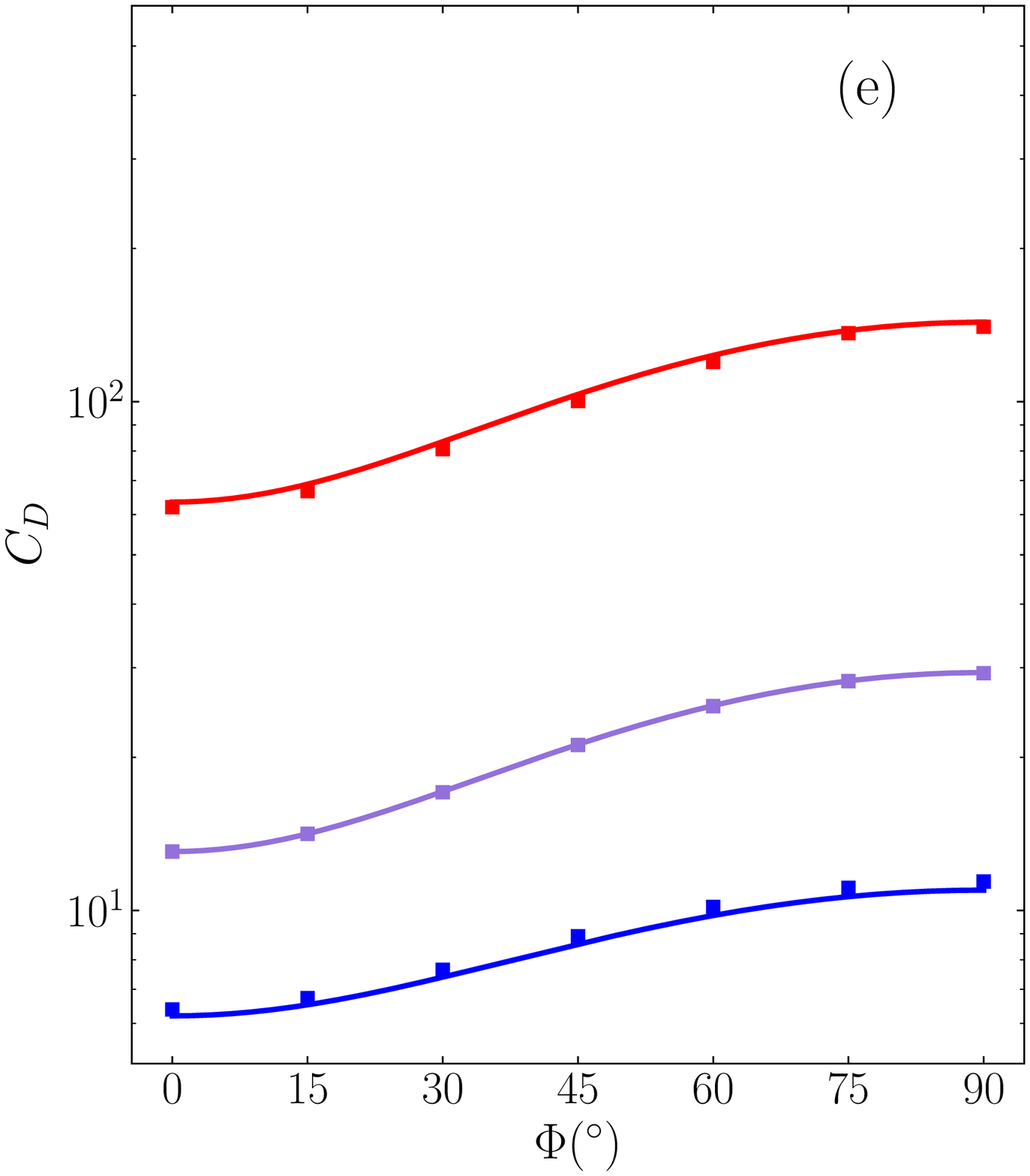}
\includegraphics[width=0.2\textwidth]{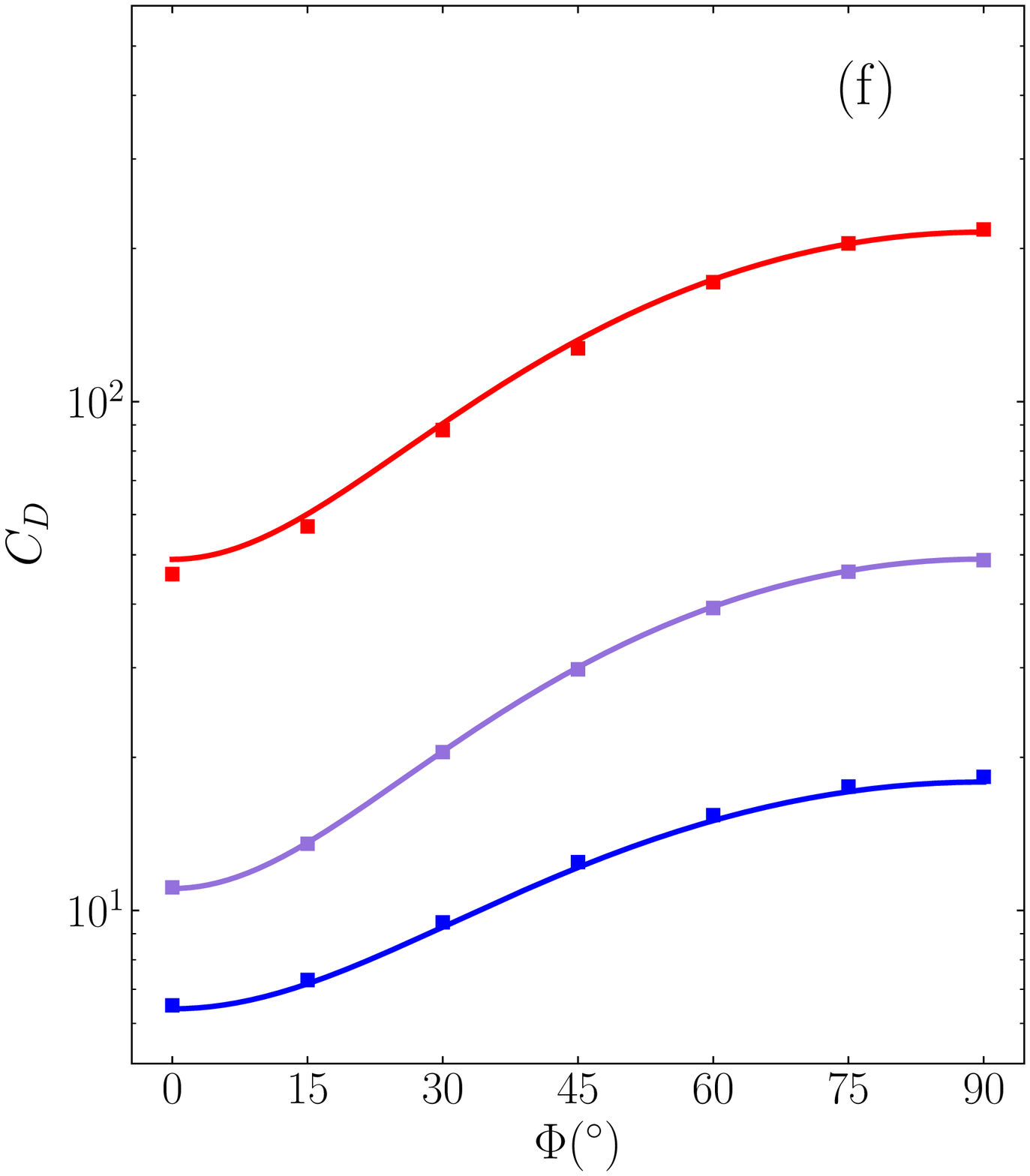}
\includegraphics[width=0.2\textwidth]{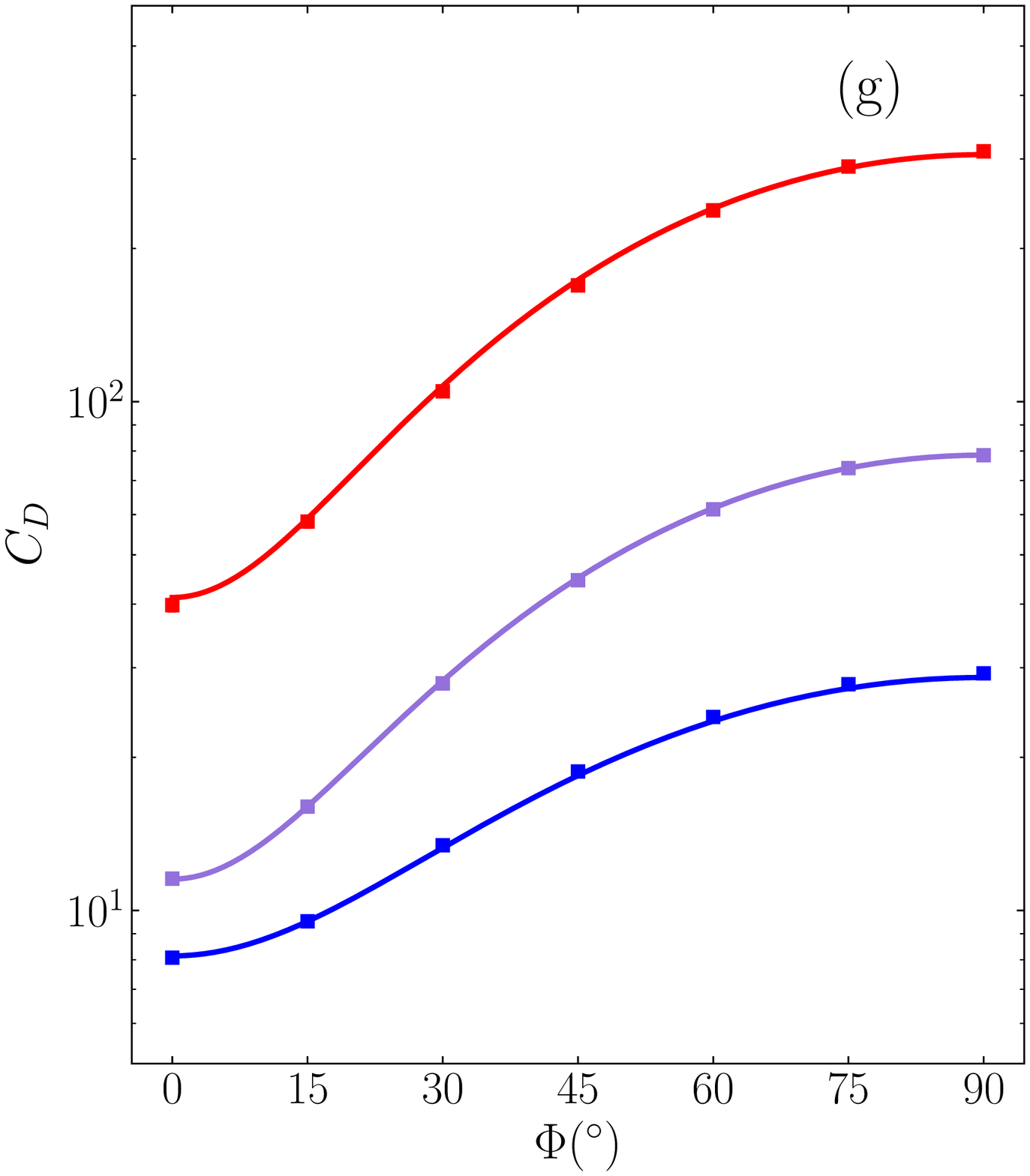}
\includegraphics[width=0.2\textwidth]{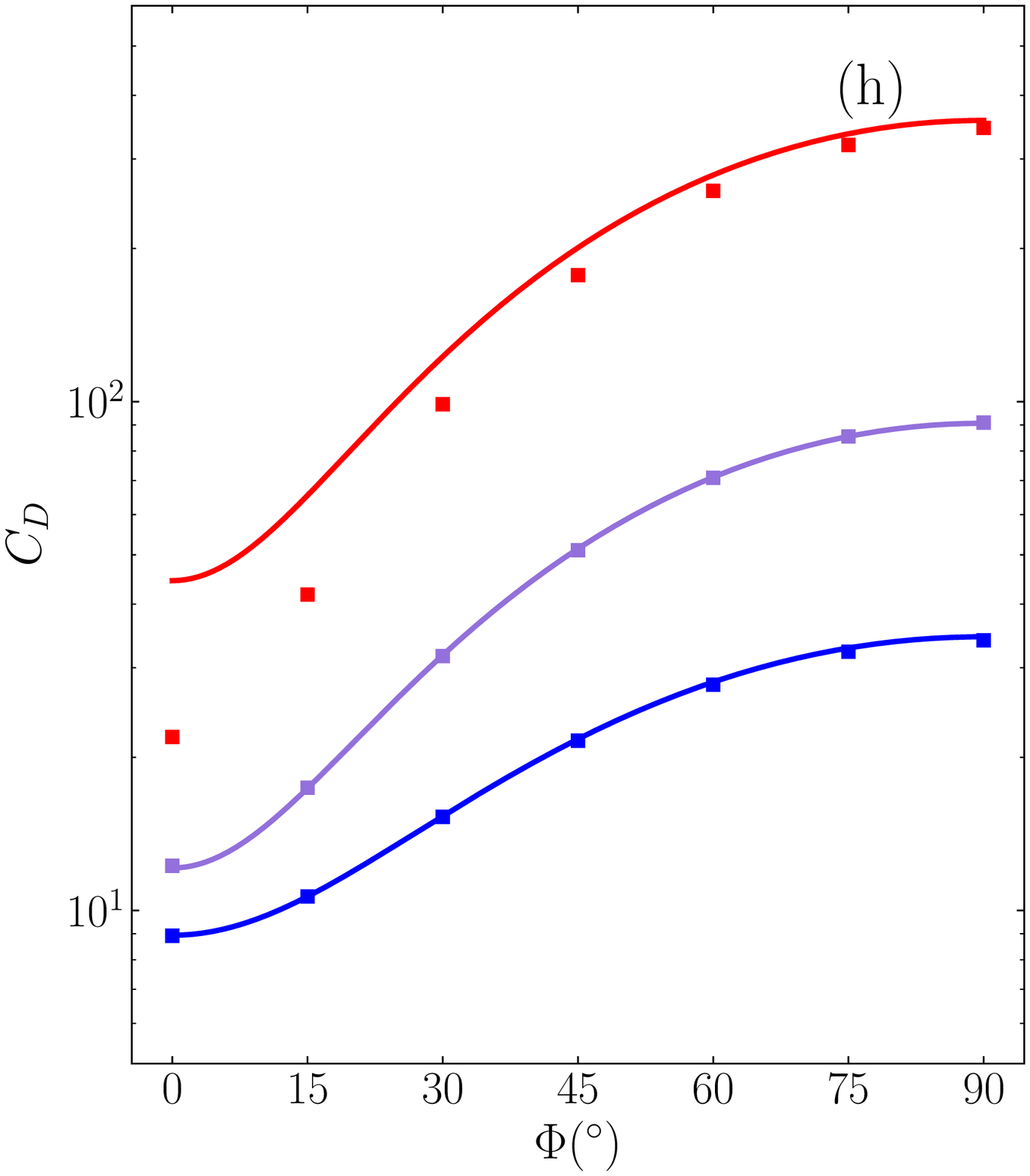}

\caption{\small{Comparison between DSMC simulations (squares) and model predictions (solid lines) of the drag coefficient $C_D$ of a prolate (a, b, c, d) and an oblate (e, f, g, h) ellipsoid, for $a/b=2$ (a,e), $a/b=4$ (b, f), $a/b=8$ (c, g) and $a/b=10$ (d, h). Every set of test has been performed with different values of $Kn$ and $\sigma$ that were not used during the fitting process.  The match between the model predictions and DSMC data is excellent for most cases, except for the oblate case at $a/b=10$ and $\sigma=0.125$. The legend is shown only for (a) as all the symbols refer to the same quantities in all plots.}}
\label{fig:test_tmac}
\end{figure*}
\begin{figure}
\centering

\begin{tikzpicture}[>=stealth,declare function={R = 0.75cm;},declare function={G = 0.4cm;}]

\node[anchor=south west,inner sep=0] (image) at (0,0) {\includegraphics[width=0.45\textwidth]{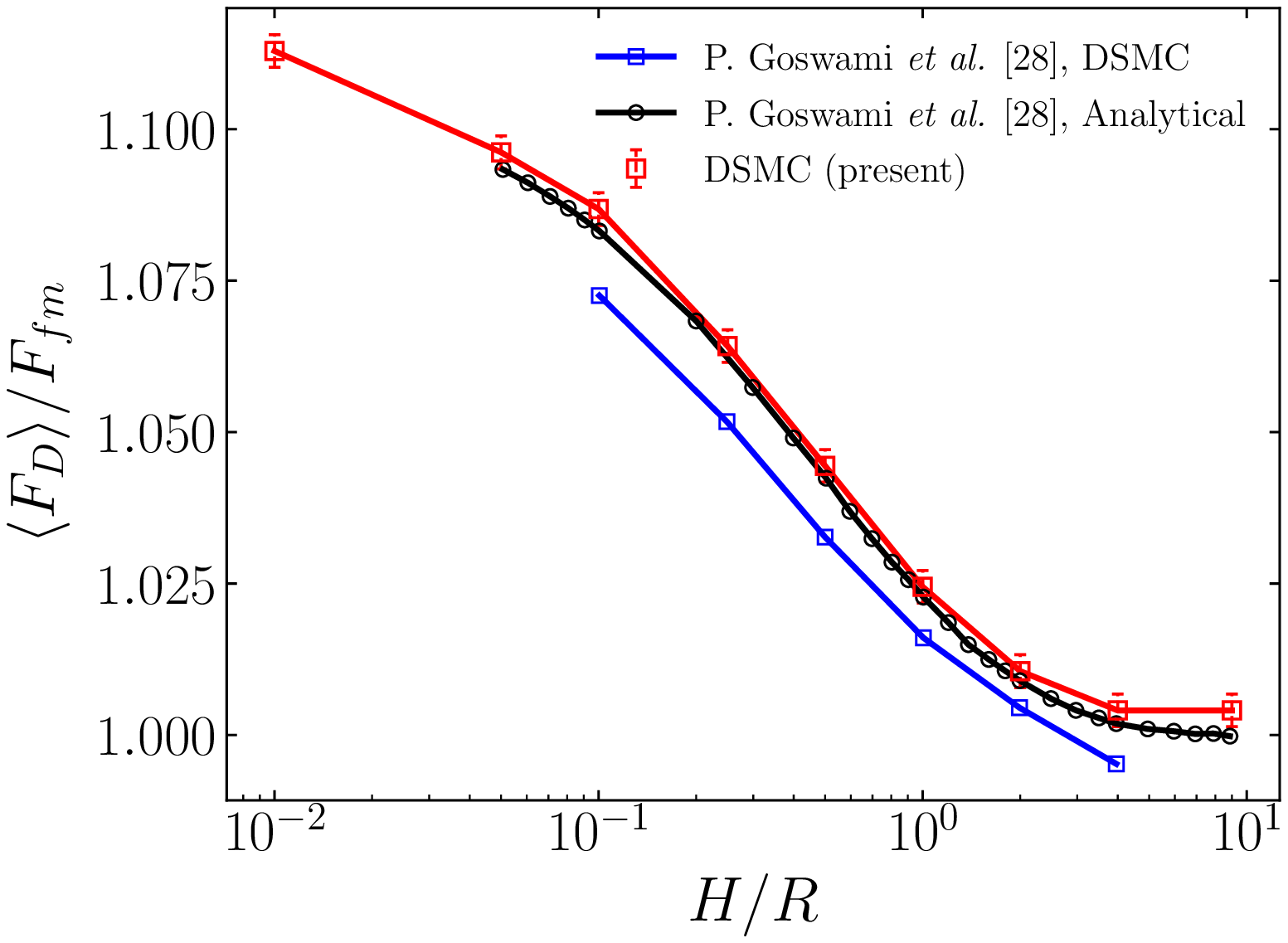}};
        
\begin{scope}[shift={(3.2cm,2.3cm)}, myVeryThick/.style={line width=1.2pt}]
        
\draw [line width=1.2pt] (0,0) circle (R);
 \fill[gray!40] (0,0) circle (R);
          
\draw[line width=1.1pt,-] (-R-1,R+G) -- (R+1,R+G);
\fill[gray!60] (-R-1,R+G) rectangle (R+1,R+G+2);
\draw [line width=0.8pt,<->] (0,R) -- (0,R+G);

\node[anchor=west,inner sep=0] at (0.1,R+0.5*G) {\small{$H$}};
\node[anchor=south west,inner sep=0] at (R*0.7,R*0.7) {\small{$R$}};
\draw [line width=0.8pt,->] (0,0) -- (R*0.7,R*0.7);
\draw [line width=0.8pt,->] (-1.2*R,0) -- (-2.1*R,0);
\node[anchor=south,inner sep=0] at (-1.6*R,0.1) {\small{$-U_0$}};

\end{scope}
\end{tikzpicture}\\
\caption{\small{Drag force experienced by a spherical particle with radius $R=0.25\mu$m translating with uniform velocity $-U_0$ parallel to a wall, for different values of the distance from the wall expressed by $H/R$. Results from DSMC simulations (red) are compared with the analytical results obtained from \cite{goswami} and their DSMC method (blue). All solid surfaces are considered fully diffusive ($\sigma=1$).}}
\label{fig:sphere_wall}
\end{figure}
In the following of this Section we show that the presented predictive model can be efficiently extended to particles with larger aspect ratio, approximating more complex shapes such as needles (for the prolate case) and flakes (for the oblate case). This aspect highlights that the perturbative approach employed to derive the model functions from Eq. (\ref{eq:g}) is not limiting the applicability of such method, even when the shape of the particle under investigation largely deviates from the spherical case.\\
By repeating the procedure explained for the case with aspect ratio $a/b=2$, we obtain the free parameters $p_\chi,\ q_\chi,\ r_\chi$ and $s_\chi$, with $\chi=0^\circ$ and $90^\circ$, for the cases with $a/b=4,\ a/b=8$ and $a/b=10$. This is, again, done by fitting the corresponding data from DSMC simulations related to the $Kn_{fit}$ set. In Fig. \ref{fig:drag_ab} we show the model functions, as well as their testing by direct comparison with DSMC results from the $Kn_{test}$ set, which have not been used during the fitting process. We focus on the cases at $0^\circ$ and $90^\circ$, since it has been shown in \cite{livi} that it is sufficient to predict the drag force at such orientations to be able to extend the prediction to any arbitrary vale of $\Phi$. Due to the large number of costly DSMC simulations required for this study, and considering that we are mostly interested in applications at $Kn\geq 1$, the test set is limited to $Kn_{test}=1,3,7,9,20$ for the following validations. The results shown in Fig. \ref{fig:drag_ab} clearly indicate that our model functions work properly for all aspect ratios considered.\\
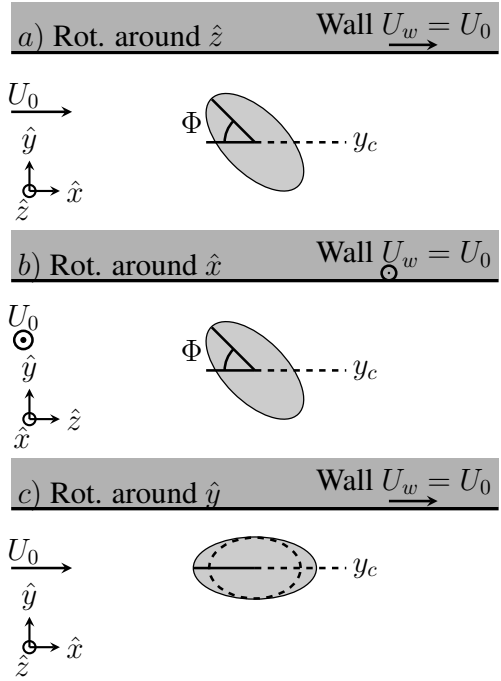
\begin{figure}[h!]
\vspace{0.2cm}

\centering
\scalebox{0.81}{
\vspace{0.1cm}

\begin{tikzpicture}[>=stealth]

    % Draw ellipse
      \begin{scope}[shift={(0,0.5cm)}, rotate around={-45:(0,0cm)},myVeryThick/.style={line width=1.2pt}]
           \draw [myVeryThick] (0,0) ellipse (1cm and 0.5 cm);
           \fill[gray!40] (0,0) ellipse (1cm and 0.5 cm);
           \draw[myVeryThick,-] (0,0) -- (-1,0cm);
      \end{scope}
      
    % angle of attack  
       \begin{scope}[shift={(0,0.5cm)}, myVeryThick/.style={line width=1.2pt}]
       	 \draw[myVeryThick,-] (0,0) -- (-0.8cm,0);
     	 \draw[color=black,myVeryThick] (-0.35cm,0.35cm) arc (135:180:0.5cm);   
         \node[anchor=east,inner sep=0] at (-0.85cm,0.32cm) {\Large{$\Phi$}};
     	 \draw[color=black,myVeryThick,dashed] (0,0) -- (1.5cm,0cm);  
         \node[anchor=west,inner sep=0] at (1.6cm,0cm) {\Large{$y_c$}};
     	 \draw[color=black,myVeryThick,->] (-4,0.5) -- (-3cm,0.5cm);  
         \node[anchor=south,inner sep=0] at (-3.8cm,0.55cm) {\Large{$U_0$}};         
         
       \end{scope}

    % wall
       \begin{scope}[myVeryThick/.style={line width=3pt}]
      	 \draw[myVeryThick,-] (-4cm,2cm) -- (4cm,2cm);
      	  \fill[gray!60] (-4cm,2cm) rectangle (4cm,2.8cm);
         \node[anchor=south west,inner sep=0] at (1,2.2) {\Large{$\mbox{Wall }U_w=U_0$}};
      	 \draw[line width=1.2pt,->] (2.2cm,2.1cm) -- (3cm,2.1cm);
         \node[anchor=west,inner sep=0] at (-3.9,2.2) {\Large{$a)$ Rot. around $\hat{z}$}};

       \end{scope}

    % axes annotations 
       \begin{scope}[shift={(-3.7cm,-0.3cm)},myVeryThick/.style={line width=1pt}]
      	       	 \draw[myVeryThick,->] (0,0) -- (0.5cm,0);
      	       	 \draw[myVeryThick,->] (0,0) -- (0,0.5cm);
      	       	 \draw[myVeryThick,->] (0,0) circle (0.1);
                \node[anchor=west,inner sep=0]   at (0.6,0) {\Large{$\hat{x}$}};
                \node[anchor=south,inner sep=0] at (0,0.6) {\Large{$\hat{y}$}};
                \node[anchor=north,inner sep=0] at (-0.12,-0.12) {\Large{$\hat{z}$}};

       \end{scope}

\end{tikzpicture}
}\\
\vspace{0.1cm}
\scalebox{0.81}{
\begin{tikzpicture}[>=stealth]

    % Draw ellipse
      \begin{scope}[shift={(0,0.5cm)}, rotate around={-45:(0,0cm)},myVeryThick/.style={line width=1.2pt}]
           \draw [myVeryThick] (0,0) ellipse (1cm and 0.5 cm);
           \fill[gray!40] (0,0) ellipse (1cm and 0.5 cm);
           \draw[myVeryThick,-] (0,0) -- (-1,0cm);
      \end{scope}
      
    % angle of attack  
       \begin{scope}[shift={(0,0.5cm)}, myVeryThick/.style={line width=1.2pt}]
       	 \draw[myVeryThick,-] (0,0) -- (-0.8cm,0);
     	 \draw[color=black,myVeryThick] (-0.35cm,0.35cm) arc (135:180:0.5cm);   
         \node[anchor=east,inner sep=0] at (-0.85cm,0.32cm) {\Large{$\Phi$}};
     	 \draw[color=black,myVeryThick,dashed] (0,0) -- (1.5cm,0cm);  
         \node[anchor=west,inner sep=0] at (1.6cm,0cm) {\Large{$y_c$}};
     	 \draw[line width=1.2pt,->] (-3.8cm,0.5cm) circle (0.15);
      	  \fill[line width=1.2pt,->] (-3.8cm,0.5cm) circle (0.05);
      	           \node[anchor=south,inner sep=0] at (-3.8cm,0.65cm) {\Large{$U_0$}};         
         
       \end{scope}

    % wall
       \begin{scope}[myVeryThick/.style={line width=3pt}]
      	 \draw[myVeryThick,-] (-4cm,2cm) -- (4cm,2cm);
      	  \fill[gray!60] (-4cm,2cm) rectangle (4cm,2.8cm);
         \node[anchor=south west,inner sep=0] at (1,2.2) {\Large{$\mbox{Wall }U_w=U_0$}};
      	 \draw[line width=1.2pt,->] (2.2cm,2.1cm) circle (0.12);
      	  \fill[line width=1.2pt,->] (2.2cm,2.1cm) circle (0.02);

         \node[anchor=west,inner sep=0] at (-3.9,2.2) {\Large{$b)$ Rot. around $\hat{x}$}};

       \end{scope}

    % axes annotations 
       \begin{scope}[shift={(-3.7cm,-0.3cm)},myVeryThick/.style={line width=1pt}]
      	       	 \draw[myVeryThick,->] (0,0) -- (0.5cm,0);
      	       	 \draw[myVeryThick,->] (0,0) -- (0,0.5cm);
      	       	 \draw[myVeryThick,->] (0,0) circle (0.1);
                \node[anchor=west,inner sep=0]   at (0.6,0) {\Large{$\hat{z}$}};
                \node[anchor=south,inner sep=0] at (0,0.6) {\Large{$\hat{y}$}};
                \node[anchor=north,inner sep=0] at (-0.12,-0.12) {\Large{$\hat{x}$}};

       \end{scope}

\end{tikzpicture}
}\\
\vspace{0.1cm}
\scalebox{0.81}{
\begin{tikzpicture}[>=stealth]

    % Draw ellipse
      \begin{scope}[shift={(0,1cm)}, rotate around={0:(0,0cm)},myVeryThick/.style={line width=1.2pt}]
           \draw [myVeryThick] (0,0) ellipse (1cm and 0.5 cm);
           \fill[gray!40] (0,0) ellipse (1cm and 0.5 cm);
           \draw [myVeryThick,dashed] (0,0) ellipse (0.75cm and 0.5 cm);
           \draw[myVeryThick,-] (0,0) -- (-1,0cm);
      \end{scope}
      
    % angle of attack  
       \begin{scope}[shift={(0,1cm)}, myVeryThick/.style={line width=1.2pt}]
            	 \draw[color=black,myVeryThick,dashed] (0,0) -- (1.5cm,0cm);  
         \node[anchor=west,inner sep=0] at (1.6cm,0cm) {\Large{$y_c$}};
     	 \draw[color=black,myVeryThick,->] (-4,0) -- (-3cm,0cm);  
         \node[anchor=south,inner sep=0] at (-3.8cm,0.05cm) {\Large{$U_0$}};         
         
       \end{scope}
    % wall
       \begin{scope}[myVeryThick/.style={line width=3pt}]
      	 \draw[myVeryThick,-] (-4cm,2cm) -- (4cm,2cm);
      	  \fill[gray!60] (-4cm,2cm) rectangle (4cm,2.8cm);
         \node[anchor=south west,inner sep=0] at (1,2.2) {\Large{$\mbox{Wall }U_w=U_0$}};
      	 \draw[line width=1.2pt,->] (2.2cm,2.1cm) -- (3cm,2.1cm);
         \node[anchor=west,inner sep=0] at (-3.9,2.2) {\Large{$c)$ Rot. around $\hat{y}$}};

       \end{scope}

    % axes annotations 
       \begin{scope}[shift={(-3.7cm,-0.3cm)},myVeryThick/.style={line width=1pt}]
      	       	 \draw[myVeryThick,->] (0,0) -- (0.5cm,0);
      	       	 \draw[myVeryThick,->] (0,0) -- (0,0.5cm);
      	       	 \draw[myVeryThick,->] (0,0) circle (0.1);
                \node[anchor=west,inner sep=0]   at (0.6,0) {\Large{$\hat{x}$}};
                \node[anchor=south,inner sep=0] at (0,0.6) {\Large{$\hat{y}$}};
                \node[anchor=north,inner sep=0] at (-0.12,-0.12) {\Large{$\hat{z}$}};

       \end{scope}

\end{tikzpicture}
}

\caption{\small{Sketch of the simulation setup: a prolate ellipsoidal particle with aspect ratio $a/b=2$ is immersed in a uniform flow with ambient velocity $U_0$. The center of the particle, $y_c$, is located in proximity to a solid, fully diffusive ($\sigma=1$) wall translating at a velocity $U_w=U_0$. In this way the problem is analogous at the one of a particle translating with uniform velocity $-U_0$ parallel to the wall. We investigate orientation effects with respect to rotations around the three main Cartesian axes $\hat{z}$ (a), $\hat{x}$ (b) and $\hat{y}$ (c), for different values of $y_c$.} Note that in case (b) the major radius $a$ is always orthogonal to the flow direction. Free stream and periodic boundary conditions are applied on the directions orthogonal to $\hat{x}$ and $\hat{z}$, respectively.}
\label{fig:ell_wall_sketch}
\end{figure}
\begin{figure*}
\centering
\includegraphics[width=0.31\textwidth]{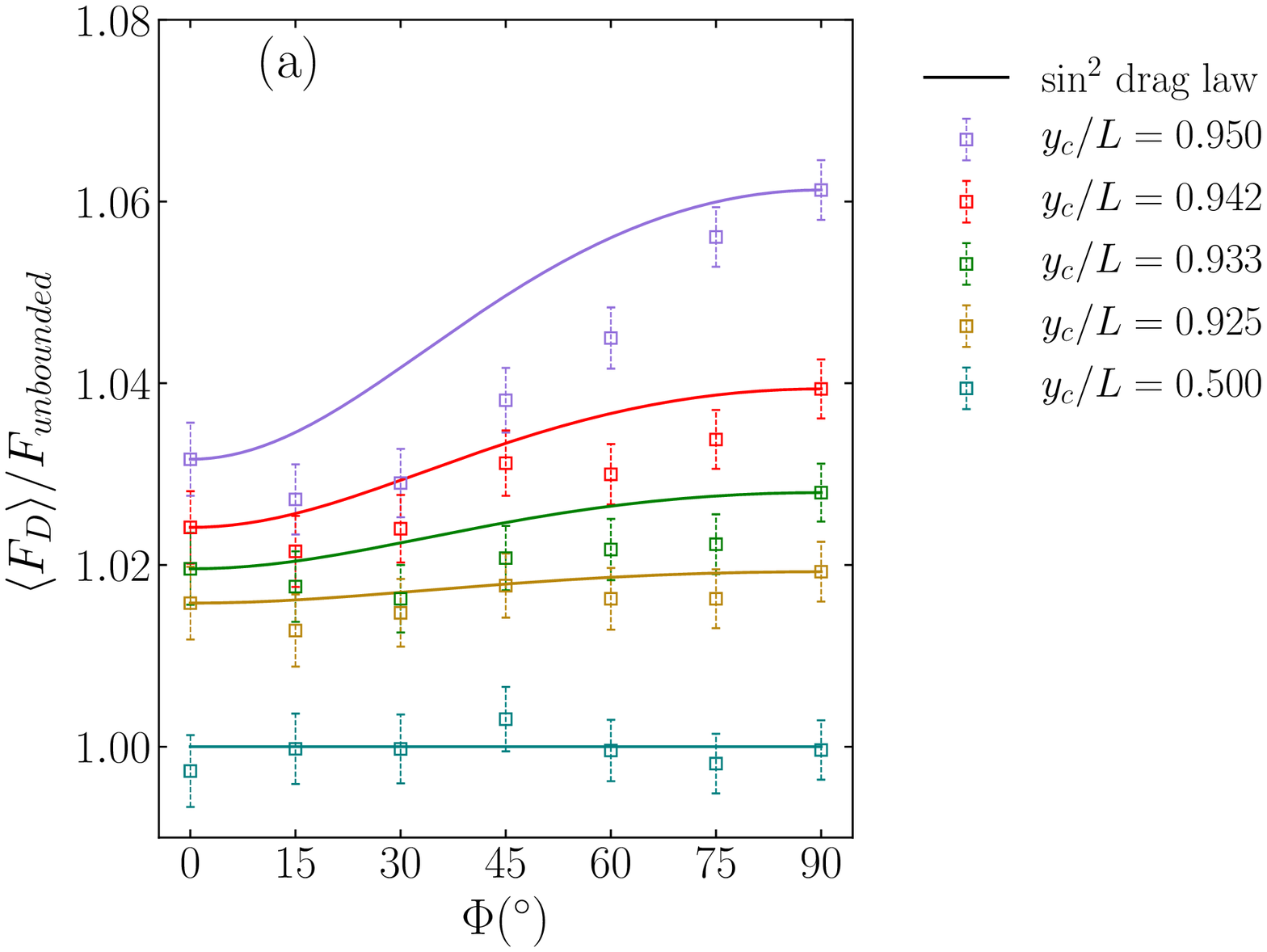}%
\includegraphics[width=0.31\textwidth]{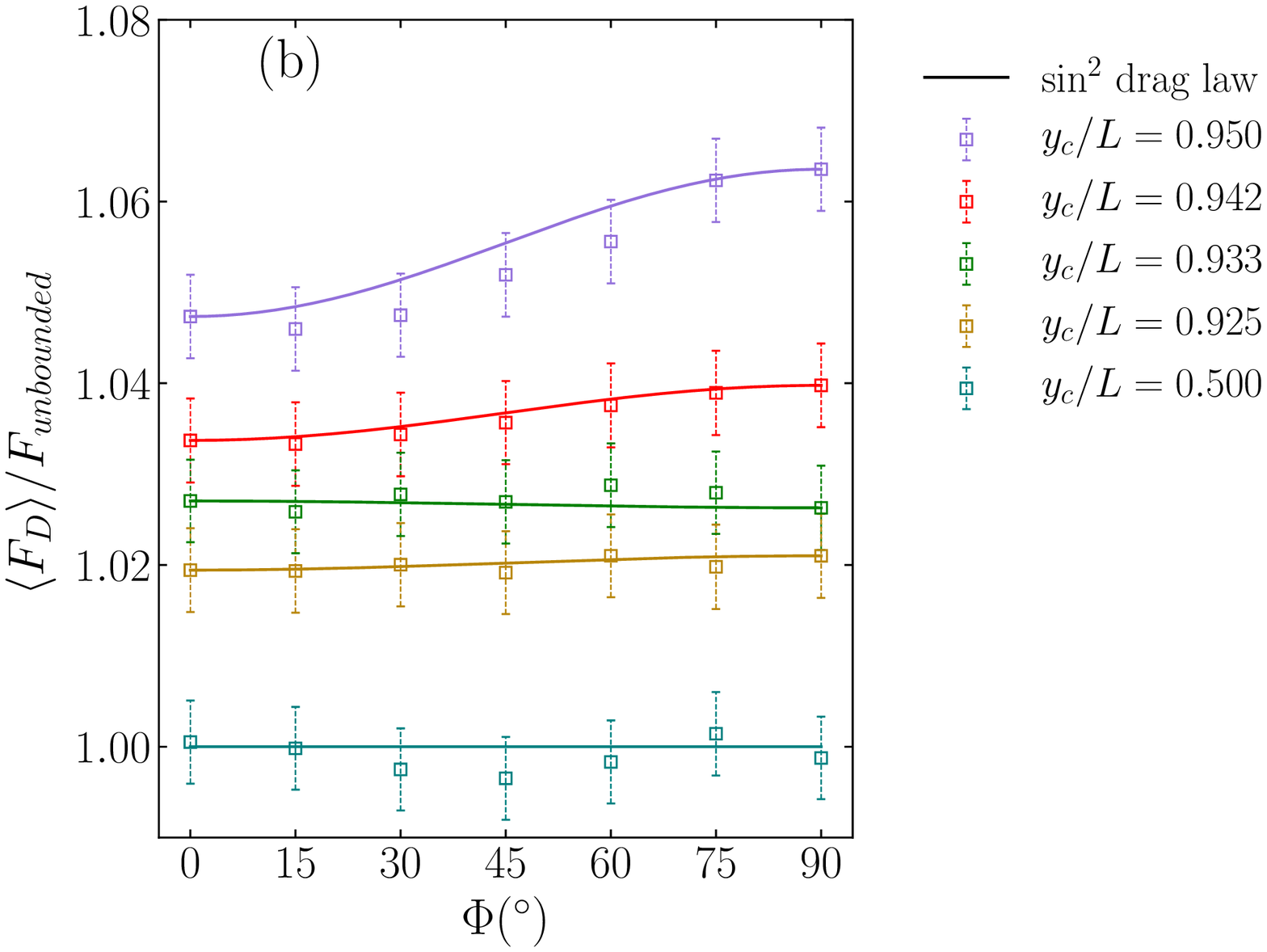}%
\includegraphics[width=0.31\textwidth]{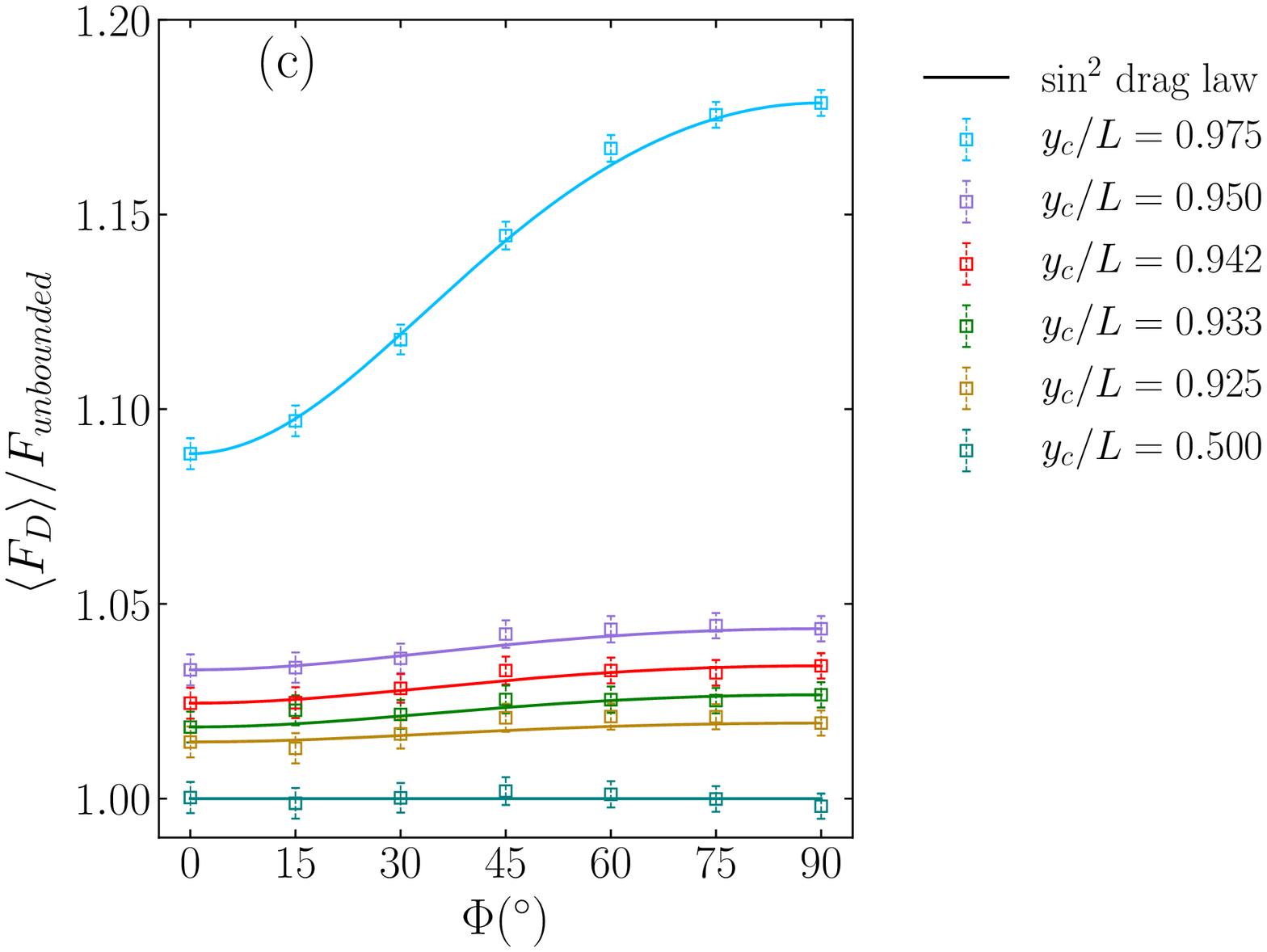}
\caption{\small{Drag force experienced by an ellipsoidal particle with aspect ratio $a/b=2$ translating with uniform velocity parallel to a wall at fixed $Kn=10$, for different vertical locations $y_c/L$, where $y_c$ is the vertical coordinate of the center of the particle and $L$ is the dimension of the simulation domain. The drag force is evaluated at different orientations $\Phi$ with respect to the ambient flow, for rotations around the $\hat{z}$ axis (a), $\hat{x}$ axis (b) and $\hat{y}$ axis (c). Results from DSMC simulations (squares) are compared with the sine-squared drag law (solid lines) from Eq. (\ref{eq:sine_squared_f}) calculated using values from DSMC simulations at $\Phi=0^\circ$ and $\Phi=90^\circ$. It is possible to observe that  Eq. (\ref{eq:sine_squared_f}) still captures the general scaling of the drag force, with the exception of case (a) and particle in close contact with the wall for which a larger deviation is evident. The error bars in this plot are obtained via $\varepsilon_{95}$ and considering the error propagation related to the drag force ratio.}}
\label{fig:ell_wall}
\end{figure*}
\begin{figure*}
\centering
\includegraphics[width=0.32\textwidth]{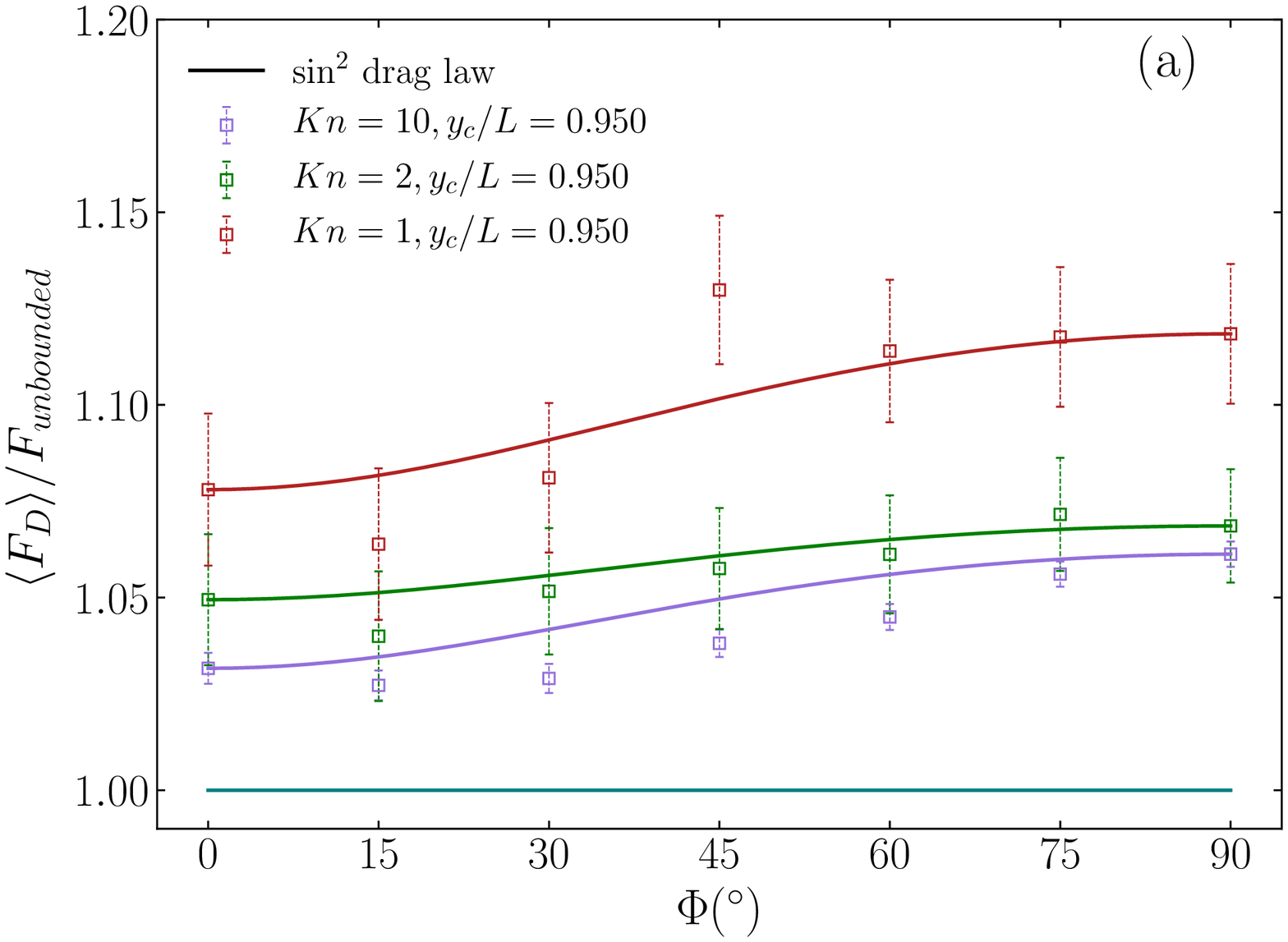}%
\includegraphics[width=0.32\textwidth]{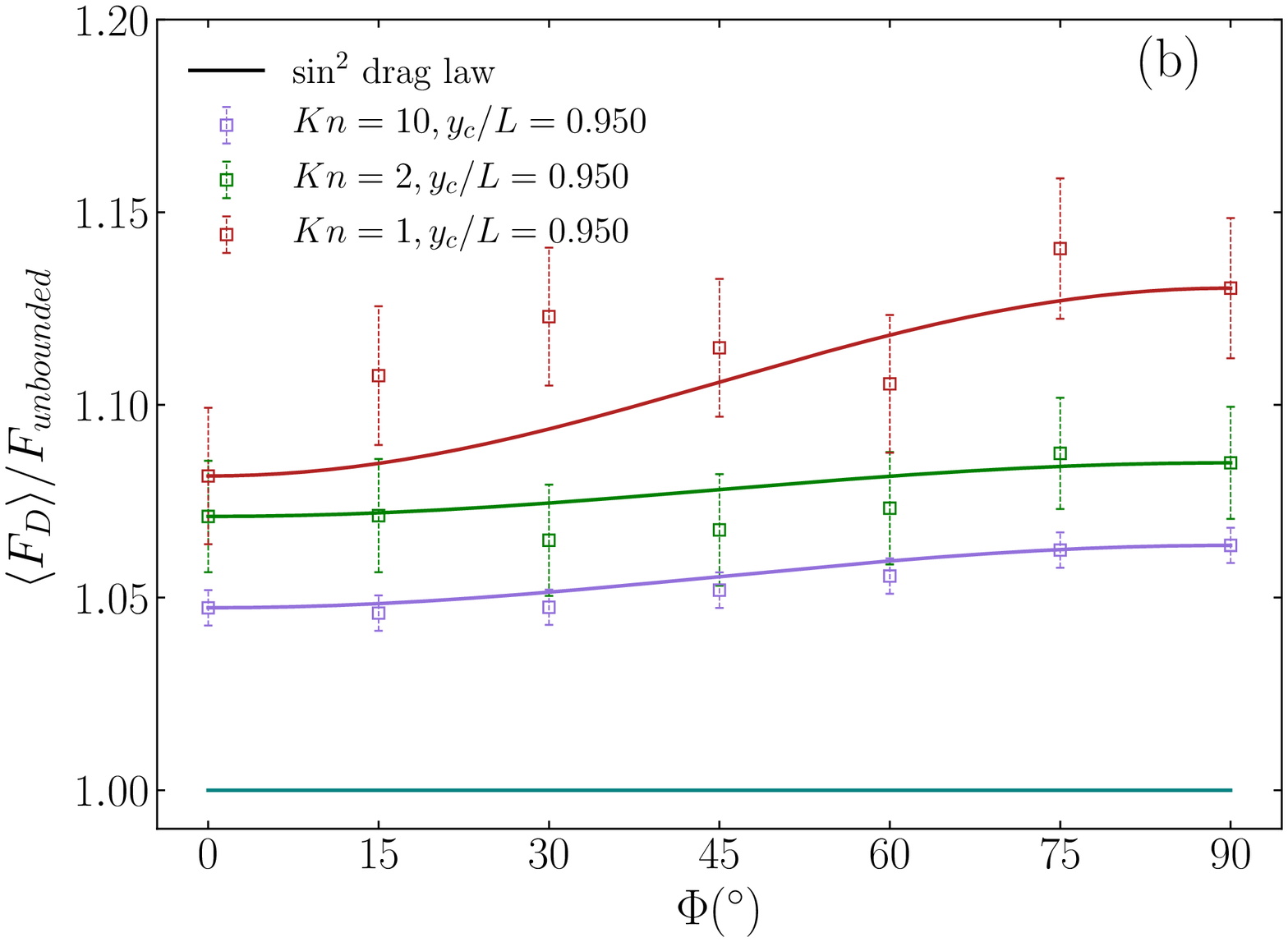}%
\includegraphics[width=0.32\textwidth]{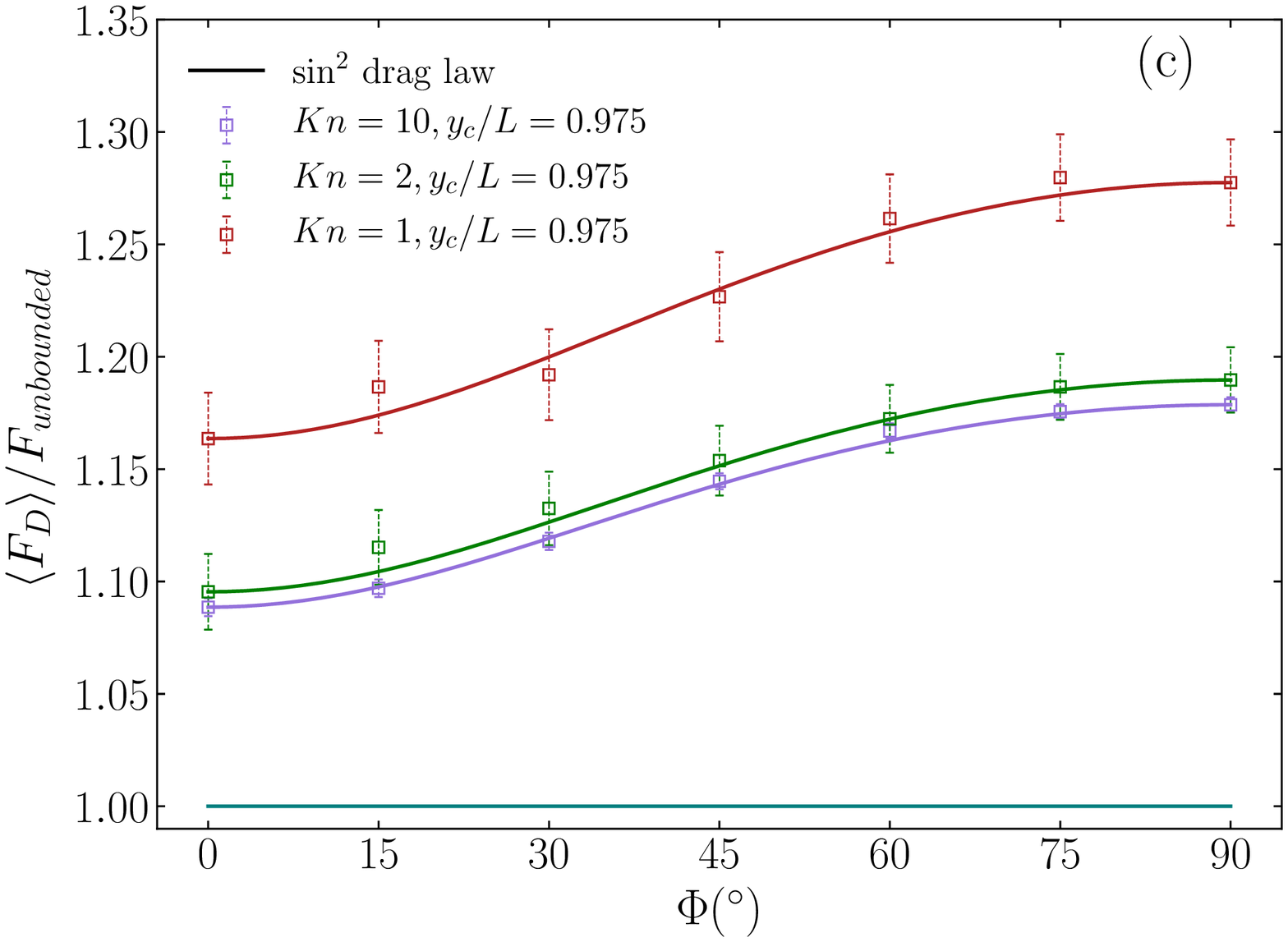}\\
\caption{\small{Drag force experienced by an ellipsoidal particle with aspect ratio $a/b=2$ translating with uniform velocity parallel to a wall for $Kn=10$ (purple), $Kn=2$ (green) and $Kn=1$ (red). The drag force, normalized with respect to the unbounded case, is evaluated at different orientations $\Phi$ with respect to the ambient flow, for rotations around the $\hat{z}$ axis (a), $\hat{x}$ axis (b) and $\hat{y}$ axis (c), while the vertical position $y_c/L$ is fixed so that the particle has one point of contact with the wall for $\Phi=90^\circ$ in cases (a) and (b)( $y_c/L=0.95$), and is always on contact for case (c) ($y_c/L=0.975$). Results from DSMC simulations (squares) are compared with the sine-squared drag law (solid lines) from Eq. (\ref{eq:sine_squared_f}) calculated using values from DSMC simulations at $\Phi=0^\circ$ and $\Phi=90^\circ$.}}
\label{fig:ell_wall_kn}
\end{figure*}
In the last part of this Section we will extend our results, so far limited to fully diffusive reflections at the surface of the solid particles, to include TMAC effects. For nano-metric particles, in fact, it is known \cite{wang} that an increasingly larger fraction of gas molecules undergoes specular reflection when hitting the surface of the solid particle, and it is important to capture such effects to correctly model particle transport in contamination control applications. We show that by adding an extra term to the model Eq. (\ref{eq:g}) we can successfully include effects of a varying TMAC, $\sigma$, in the prediction of the drag coefficient. The modified model functions are defined as:
\begin{align}
g'_{\chi}(Kn,\sigma) = g_{\chi}(Kn)+ \frac{\alpha_{\chi}Kn}{\beta_{\chi} + \gamma_{\chi}Kn^{\delta_{\chi}}}(1-\sigma),
\label{eq:g_tmac}
\end{align}
where an extra term containing the $\sigma$-dependence is included in the previously defined model functions (represented by the first term in the right end side). This new term is used to model the corrections to the drag given by the presence of a combination of specular and diffuse reflection at the solid interface, with $0\leq \sigma \leq 1$, and it is designed to vanish for $\sigma=1$, recovering $g_{\chi}(Kn)$. The extra free parameters $\alpha_\chi,\ \beta_\chi,\ \gamma_\chi$ and $\delta_\chi$ are to be determined through a 2D fit in the independent variables $Kn$ and $\sigma$, where again $\chi$ refers to the two cases at $\Phi=0^\circ$ or $Phi=90^\circ$, separately. The results from the fit for the case of a prolate ellipsoid with $a/b=2$ are presented in Fig. \ref{fig:drag_tmac}. We repeat the fitting procedure for different aspect ratios and accommodation coefficients. The testing is performed on values of $Kn$ and $\sigma$ that were not used during the fitting process, to evaluate the predictive capability of the model obtained with this procedure. The results of the testing are presented in Fig. \ref{fig:test_tmac}, where it can be observed that the predictive model is able to recover DSMC data with fairly good accuracy for all investigated cases. The final fit coefficients derived in this Section are reported in Table \ref{tab:fit}.\\
\begin{table}
\vspace{0.2cm}
\begin{center}
\begin{tabular}{ |c|c|c|c|c|c|c|c|c|c| } 
%\begin{tabular}{ |p{0.5cm}|p{1cm}|p{0.5cm}|p{0.5cm}|p{0.5cm}|p{0.5cm}|p{0.5cm}|p{0.5cm}|p{0.5cm}|p{0.5cm}| } 

\hline
 \multicolumn{10}{|c|}{   Prolate $\Phi=0^\circ$}  \\[0.1pt]
\hline
$a/b$ &$C_{D,cont}$ &$p$ & $q$ &$r$ & $s$ &$\alpha$ & $\beta$ &$\gamma$ & $\delta$ \\[0.1pt]
\toprule
$2$     & $236$         &  $-0.378$ &  $1.26$   &  $4.30$  &   $1.90$ & $0.299$ & $ 0.436$ & $ 1.08\ $ & $ 1.98$   \\ 
$4$     & $242$      & $-0.133$  &  $0.64$   &  $1.05$  & $ 1.99$ & $0.433$ & $ 0.253$ & $ 1.24\ $ & $ 1.95$ \\ 
$8$     & $278$      & $-0.122$  &  $ 0.167$ &  $  1.16$ &  $  1.90$ & $0.443$ & $ 0.323$ & $ 1.06\ $ & $ 1.98$  \\ 
$10$   & $295$       & $-0.112$  &  $ 0.218$  &  $  0.987$  &  $  1.94$ & $0.430$ & $0.294$ & $  1.04\ $ & $ 1.97$   \\ 

\hline
\end{tabular}\\
\begin{tabular}{ |c|c|c|c|c|c|c|c|c|c| } 
\hline
 \multicolumn{10}{|c|}{   Prolate $\Phi=90^\circ$}  \\[0.1pt]
\hline
$a/b$ &$C_{D,cont}$ &$p$ & $q$ &$r$ & $s$ &$\alpha$ & $\beta$ &$\gamma$ & $\delta$ \\[0.1pt]
\toprule
$2$     & $270$         & $0.094$ & $3.90$ & $1.52$ & $1.99$ & $0.151$ & $ 1.73$ & $ 1.16$ & $ 2.04$ \\ 
$4$     & $311$      & $0.131$ & $0.979$ & $1.36$ & $1.90$ & $0.125$ & $ 1.98$ & $ 0.813$ & $ 2.21$ \\ 
$8$     & $391$      & $0.122$ & $ 1.79$ & $ 1.10$ & $ 1.94 $ & $0.119$ & $ 1.99$ & $ 1.24$ & $ 2.04$\\ 
$10$   & $425$       & $0.086$ & $1.554$ & $ 0.854$ & $ 1.92$ & $0.093$ & $ 0.994$ & $ 1.27$ & $ 1.93$  \\ 

\hline
\end{tabular}\\

\begin{tabular}{ |c|c|c|c|c|c|c|c|c|c| } 
\hline
 \multicolumn{10}{|c|}{   Oblate$\Phi=0^\circ$}  \\[0.1pt]
\hline
$a/b$ &$C_{D,cont}$ &$p$ & $q$ &$r$ & $s$ &$\alpha$ & $\beta$ &$\gamma$ & $\delta$ \\[0.1pt]
\toprule
$2$     & $247$ & $-0.118\ $ & $  1.84$ & $  2.29$ & $   1.88$ & $ 0.308$ & $ 0.267$ & $ 1.29\ $ & $ 1.99$   \\ 
$4$     & $260$ & $-0.064\ $ & $  0.024$ & $  1.32$ & $  1.90$ & $ 0.436$ & $ 0.441 $ & $1.27\ $ & $ 1.96$ \\ 
$8$     & $300$ & $0.010\ $ & $ 1.01$ & $   1.99$ & $   2.10$ & $  0.547$ & $ 0.328$ & $ 1.15\ $ & $1.96$\\ 
$10$   & $318$ & $0.057\ $ & $  7.62$ & $ 1.79$ & $ 2.01$ & $  0.591$ & $ 0.447 $ & $1.15\ $ & $ 1.96$ \\ 

\hline
\end{tabular}\\
\begin{tabular}{ |c|c|c|c|c|c|c|c|c|c| } 
\hline
 \multicolumn{10}{|c|}{   Oblate $\Phi=90^\circ$}  \\[0.1pt]
\hline
$a/b$ &$C_{D,cont}$ &$p$ & $q$ &$r$ & $s$ &$\alpha$ & $\beta$ &$\gamma$ & $\delta$ \\[0.1pt]
\toprule
$2$     & $282$ & $0.307\ $ & $  3.01$ & $ 1.76$ & $ 2.00$ & $ 0.040$ & $ 1.99$ & $ 0.336$ & $ 2.52$ \\ 
$4$     & $330$ & $0.442\ $ & $ 1.29$ & $ 1.22$ & $ 1.91$ & $-0.046$ & $  1.01$ & $  1.37$ & $  1.71$ \\ 
$8$     & $410$ & $0.646\ $ & $ 1.42$ & $ 1.12$ & $1.90$ & $ -0.096$ & $  1.33$ & $  1.24 $ & $ 1.74$\\ 
$10$   & $441$ & $0.748\ $ & $ 0.840$ & $ 1.44 $ & $1.80$ & $ -0.055 $ & $ 2.00$ & $  0.136$ & $  2.42$ \\ 

\hline
\end{tabular}
\end{center}
\caption{\small{Fit parameters obtained using Eq. (\ref{eq:ellipsoid_prediction}) to fit $C_{D,0^\circ}$ and $C_{D,90^\circ}$ as obtained from DSMC simulations performed at $Kn_{fit}$ for all cases investigated in this work. The parameters $p,q,r,s$ are used to define the function $g_{\chi}(Kn)$ given by Eq. (\ref{eq:g}), while the parameters $\alpha,\beta,\gamma,\delta$ are used to define the function $g'_{\chi}(Kn)$ given by Eq. (\ref{eq:g_tmac}). Additionally, the drag coefficients in the continuum regime $C_{D,cont}$, computed from \cite{oberbeck} and used as reference, are presented.}}
\label{tab:fit}
\end{table}
\section{Drag corrections from near-wall effects}
\label{sec:2}

The approach discussed in Section \ref{sec:1} is formally valid only for the case of a particle immersed in a uniform unbounded flow. In many practical situations, however, the particles transported in the flow will interact with other elements of the flow domain, such as walls or other obstacles and it is important to understand the limitations in the applicability of the methods proposed in this work when particles are located in proximity of a wall. The aim of this  Section is to demonstrate that the drag corrections derived in Section \ref{sec:1} can adopted also in non-ideal situations where the unbounded flow condition is no longer valid, provided that the particle-based $Kn$ is large and that the particle size is much smaller than the typical size of the system in order to maintain the Point-Particle approximation.\\
We tackle the problem by evaluating the impact of the presence of a wall on the drag experienced by an ellipsoidal particle translating parallel to it. This condition is easily achievable by performing simulations in the reference frame of the particle, so that it is sufficient to modify the setup discussed in Section \ref{sec:1} by adding solid walls on the direction orthogonal to $\hat{y}$ moving with the same ambient velocity of the flow (see Figs. \ref{fig:sphere_wall} and \ref{fig:ell_wall_sketch}). In this way the problem is analogous to the one of a particle translating with a velocity $-U_0$ parallel to the wall. All the simulations are performed with the same criteria discussed in Section \ref{sec:1}, with the exception that now the particle center is located close to the upper edge of the simulation box, at which a solid moving wall condition is applied. \\
To validate the simulation setup, we compute the drag acting on a sphere translating parallel to the wall in the collisionless regime and compare with the results from \cite{goswami}. The comparison is shown in Fig. \ref{fig:sphere_wall}, where it can be seen that the drag acting on the spherical particle measured from our DSMC simulations correctly reproduces the analytical expression obtained by \cite{goswami}.\\
We now proceed in investigating the effects on the drag force experienced by an ellipsoidal particle translating in the vicinity of a wall. Due to the large number of parameters, we will restrict our study to a prolate ellipsoid with aspect ratio $a/b=2$ and by assuming that all the solid surfaces (i.e. both wall and particles) are fully diffusive ($\sigma=1$). The focus on the prolate case limits our claims to this specific geometry, however we expect similar results for the oblate case and for the other aspect ratios.\\
In our first analysis, we perform simulations varying the orientation and the distance of the particle from the top wall of the simulation box, addressing separately three different cases of rotation around the $x,y$ and $z$ axes. Due to the presence of the wall, in fact, the symmetry of the system is no longer conserved and rotations around different axes are expected to produce different results. For the sake of simplicity, we focus on the rotations around the three main Cartesian axes, where for the case of rotation around the $x$ axis the ellipsoidal particle major axis is orthogonal to the flow direction (constant angle of attack). A sketch of the simulation setup is presented in Fig. \ref{fig:ell_wall_sketch}.\\
In Fig. \ref{fig:ell_wall} we compare the drag force experienced by the particle, at fixed $Kn=10$, for different orientations and different vertical locations of the particle (i.e. different distances from the wall). The drag force is normalized with respect to the drag force relative to the unbounded case to emphasize the effects introduced by the wall, while the vertical position of the center of the particle, $y_c$, is normalized with respect to the simulation box size, $L$. The largest value of $y_c/L$ is chosen so that the particle is in contact with the wall at $\Phi=90^\circ$ (with the exception of case (c), where $y_c$ is chosen so that the particle is in contact for all values of $\Phi$).\\
A deviation from the unbounded case (horizontal line in  Fig. \ref{fig:ell_wall}) is evident, as drag experienced by the particle increases the closer its surface gets to the wall. We observe that this drag-increase effect depends, as expected, not only to the vertical position, but also to the rotation angle $\Phi$ for cases (a) and (b), since for larger values of $\Phi$ the surface of the particle is closer to the solid wall, while for case (c) the orientation only impacts on the squeezing of the flow between the particle and the wall. Interestingly, we also observe that the general sine-squared scaling is maintained if the drag force evaluated at $\Phi=0^\circ$ and $\Phi=90^\circ$ are used in Eq. (\ref{eq:sine_squared_f}), and the largest variations from it appear in case (a) for particle in close contact with the wall . The drag increase induced by the plane wall quickly vanishes as the particle moves away from it, to the point that for $y_c/L=0.925$ (which corresponds to a wall distance equivalent to the particle major radius, $a$) such effects are already lower than $2\%$.\\
In the last analysis of this work, we compare the drag force experienced by the near-wall particle for three values of $Kn$, i.e. $Kn=10$ (free molecular regime), $Kn=2$ and $Kn=1$  (transitional regime). The vertical position of the particle, $y_c/L$, is chosen so that the particle is in contact with the wall at $\Phi=90^\circ$ for the cases of rotation around the $\hat{z}$ (a) and $\hat{x}$ (b) axes , while it is always in contact for the case of rotation around the $\hat{y}$ axis (c). To better resolve the physics in the gap between the particle and the solid wall and improve the signal-to-noise of DSMC simulations, we increased the resolution for $Kn<10$. For the case with $Kn=2$ it was sufficient to increase the PPC number to $150$, while for the case with $Kn=1$ we also increase the spatial resolution. In the latter case we set a linear domain size of $256$ grid cells, leading to a particle major radius $a=12.7$ (in cell units); the PPC number is set to $100$.\\
As it can be seen from the results in Fig. \ref{fig:ell_wall_kn}, the drag increase, with respect to the unbounded case, exhibits a weak dependence on the Knudsen number, especially when $Kn>1$. In the worst case scenario ($\Phi=90^\circ$ and rotation around $\hat{z}$ axis), the deviation with respect to the unbounded case is about $27\%$ for the case at $Kn=1$ and it quickly drops to $19\%$ for the case at $Kn=2$; after this, it remains roughly constant up to $Kn=10$, where we measure an increase of about $18\%$. Near-wall effects appear, thus, to be weak enough to be considered negligible for the vast majority of applications involving the transport of micro- and nano-metrical ellipsoidal particles in highly rarefied regimes, as the average residence time of particles in areas very close to the wall (on the order of the particle characteristic size) is typically very limited, allowing the drag corrections of Section \ref{sec:1} to be applied without introducing large errors also in bounded flow situations.

\section{Conclusions}
\label{sec:conc}
We propose a new formulation and derivation of heuristic predictive models for the drag coefficients of prolate and oblate ellipsoidal particles under rarefied conditions in a range of particle-based $Kn$ that includes the transition and the free-molecular regimes. The predictive models are based on a perturbative approach, where rarefaction effects on ellipsoidal particles are represented as perturbation with respect to the spherical case and such perturbations are obtained through a fit of DSMC simulation data. \\
We firstly show that the new models are designed to recover the free-molecular and continuum asymptotic limits, potentially providing a valid baseline to improve the fitting procedure and extend the model to the whole range of Knudsen. We then obtain predictive models for a wide range of particle aspect ratios (including complex shapes such as needles and flakes), and momentum accommodation effects. For nano-metric particles, in fact, a larger fraction of specular reflection can occur at the gas-particle interface and it is important to correctly capture these effects to obtain accurate predictions. The models obtained with this procedure show robust performances in reproducing DSMC data, also outside of the range of $Kn$ used for the fitting procedure.\\
The results from this work can be used to further improve the available models of particle transport in rarefied gas flows, as the full dynamics of prolate or oblate particles under the influence of the local fluid velocity field can now be included in the aforementioned in Euler-Lagrangian simulations typically employed for different modern applications, such as contamination control in high-tech mechanical systems or aerospace engineering.\\
In the last part of the paper we show that the aforementioned models, formally derived for the case of unbounded flow, can be applied also in cases where the unbounded condition is not strictly preserved. We do so by investigating the effects induced on the drag experienced by the particles by the presence of a wall, showing that while near-wall effects increase the drag force, this effects is rather limited for large Knudsen number, and it quickly decays as the particle moves away from the wall. For the worst case scenario of a particle in contact with the wall, we measured a maximum drag increase of roughly $27\%$ at $Kn=1$, and such increase sharply decrease to $19\%$ at $Kn=2$, while for a larger value of $Kn=10$, the maximum increase is observed to be about $18\%$. Considering the rapid decay of drag-increase effects due to wall proximity, such effects can be considered negligible in the high-Knudsen range typical of the aforementioned applications as the particle residence time in the near proximity of walls is on average limited.
\section*{Acknowledgments}
This work was supported by the Netherlands Organization for Scientific Research (NWO-TTW), under the Project No. 15376 at it was carried out on the Dutch national e-infrastructure with the support of SURF Cooperative, project number 2021.035.

\bibliography{references.bib}
\bibliographystyle{unsrt}
\end{document}